# Polarity-Programmed Topological Gating of Knotting in Spatially Patterned Active Polymers

Shiqi Zhou*

School of Physics, Central South University, Changsha, Hunan, China

* Author to whom correspondence should be addressed.

Electronic mail: chixiayzsq@163.com

Spatially patterned tangential activity provides a route to program polymer topology through both geometry and propulsion direction. We use Langevin-dynamics simulations of a flexible, self-avoiding linear polymer to determine how active-domain placement and independently assigned propulsion polarities control nonequilibrium knotting. Reversing the polarities of two terminal active blocks switches the steady state from a nearly always-knotted regime to effective knot suppression, changing the knotting probability by more than three orders of magnitude despite identical active-block positions and activity magnitude. Across symmetric terminal two-block architectures, we further identify an approximate zero-one-two free-end-alignment rule: zero, one, and two free-end-aligned active blocks correspond, respectively, to suppressed, intermediate, and high-knotting states. A two-state kinetic decomposition shows that the conditional knot-formation rate changes only by about threefold under polarity reversal, whereas the mean knot lifetime changes by several orders of magnitude, demonstrating that knot survival and passive-region storage dominate the polarity-controlled switch. Block placement provides an additional geometrical gate: moving an active domain from the chain interior to a terminal region strongly enhances both knot formation and survival. Knot-center residence, knotted-arc distributions, and episode-resolved kymographs reveal formation near free-end-aligned active blocks, directed contour migration, tightening, and long-lived retention in passive regions. Together, these results support a source–conveyor–reservoir–sink mechanism in which signed polarity and spatial activity patterning control whether knots are generated, transported, stored, or rapidly lost, providing a route to programmable topological states in nonequilibrium active polymers.

## 1 Introduction

Knots are unavoidable in sufficiently long self-avoiding chains, yet their probability becomes very small when the polymer contour is short. This basic competition between topology, excluded volume, and conformational entropy underlies the statistical topology of polymers[1,2,3,4,5]. In molecular and biological settings, a knot is not merely a geometrical curiosity[6,7]: it can change mechanical response, relaxation, transport, and access to chain ends[8,9,10,11,12]. The challenge is therefore not only to detect knots, but to identify physical protocols that create, move, tighten, store, or remove them on demand[7,12].

Active polymers provide such a nonequilibrium setting. In these systems, energy-consuming monomers or molecular motors generate persistent forces that are coupled to the polymer backbone[13,14,15,16]. Activity may swell, compact, soften, stretch, or transport a chain depending on the propulsion rule and the spatial distribution of driven monomers[17,18,19,20,21,22,23,24,25,26]. Recent work in active statistical mechanics has also used generalized Langevin descriptions to characterize transport and fluctuations in self-propelled nonequilibrium systems.[27] Spatially structured driving can further reorganize active-particle trajectories and spatial distributions,[28] underscoring the broader role of heterogeneous forcing in controlling nonequilibrium dynamics. Explicit motor models have even been reported to suppress knotting relative to equilibrium, emphasizing that activity does not generically enhance topology; the direction and spatial organization of force transmission matter[29].

Recent work established that active motion can be used to engineer polymer knots. An actively reptating, end-grafted chain can repeatedly self-knot and transport knots through a nonequilibrium ratchet mechanism.[30] Vatin, Orlandini, and Locatelli showed that a terminal polar active–passive diblock can exhibit an orders-of-magnitude enhancement of spontaneous knotting and attributed its nonmonotonic dependence on active fraction to competition between knot formation in the active block and knot lifetime in the passive block[31]. Their results further established formation near the active leading end, migration toward the passive domain, and long residence of localized knots in passive contour. Active forces have also been shown to localize and tighten pre-existing knots[32]. Here we formulate a distinct topological-control problem: whether knot formation, contour transport, retention, and suppression can be selected by independently programming the spatial placement and signed propulsion polarity of active domains along a polymer contour. After validation against published diblock observables, we use the numerical framework to interrogate translated single blocks, split active domains, and independently signed local propulsion.

The key question addressed here is therefore more specific than whether activity promotes knotting: Can the topology of a short active polymer be programmed by placing active blocks at selected contour positions and assigning their propulsion directions independently? This question introduces three coupled controls. First, the active-block position determines whether an active leading edge is close to a free end, an active-passive interface, or the chain interior. Second, the number of active blocks changes the number of possible nucleation zones and interfaces. Third, the signed polarity controls the direction of backbone-coupled driving and hence the direction in which conformational folds, chain ends, and existing knots are transported.

We analyze these controls using a kinetic and spatial decomposition. The steady-state knotting probability $P_K$ is separated into the conditional knot-formation rate $\alpha_K^0$ and the mean lifetime $\langle \tau_K \rangle$. Knot-center residence distinguishes active and passive storage, while the active fraction of the shortest knotted arc distinguishes where the knot center sits from which beads constitute the entire knot. Full distributions of knot position and length reveal broad, multimodal, or long-tailed populations that are hidden by averages. Finally, episode-resolved (i.e., single-event-resolved) kymographs directly test whether a knot forms near an active end, migrates across an interface, tightens in a passive region, or disappears rapidly.

The combined data support a polarity-programmed source–conveyor–reservoir–sink framework in which active-domain placement and signed polarity determine whether knot nucleation, directed contour transfer, passive retention, or rapid escape dominates. In the studied terminal geometries, a free-end-aligned active block acts as a topological source by promoting end-assisted threading; tangentially driven contour conveys a newly formed knot away from the source; passive contour provides a reservoir in which the knot localizes and survives; and reversed or unfavorably positioned active blocks interrupt the pathway as effective barriers or sinks. Within this framework, the terminal-diblock behavior corresponds to one limiting realization, whereas independently signed and spatially placed active domains permit the complete topological pathway to be opened, redirected, or closed. The resulting mechanism enables topological gating between persistent knot storage and effective knot suppression through polarity reversal, split active domains, and terminal placement.

## 2 Model and Methods

### 2.1 Coarse-grained polymer model

The polymer is modeled as a fully flexible, self-avoiding linear bead-spring chain of $N = 300$ monomers, comprising 299 sequential bonds that confirm a single open topology. Nonbonded monomers interact through a truncated and shifted Lennard-Jones potential at the Weeks-Chandler-Andersen cutoff [33],

$$U_{\mathrm{WCA}}(r) = \begin{cases} 4\epsilon\left[\left(\frac{\sigma}{r}\right)^{12} - \left(\frac{\sigma}{r}\right)^{6}\right] + \epsilon, & r \le 2^{1/6}\sigma, \\ 0, & r > 2^{1/6}\sigma. \end{cases}$$

Neighboring beads are connected by the finitely extensible nonlinear elastic potential used in the Kremer-Grest class of bead-spring models [34],

$$U_{\mathrm{FENE}}(r) =- \frac{1}{2}KR_0^2\ln \left[1 - \left(\frac{r}{R_0}\right)^2\right] + U_{\mathrm{LJ,bond}}(r).$$

Throughout this paper, we employ Lennard-Jones (LJ) units with with $m = \sigma = 1$, nonbonded energy $\epsilon = 10$, cutoff $2^{1/6}\sigma$, and FENE parameters $K = 300$, $R_0 = 1.5$, $\epsilon_{\mathrm{bond}} = 10$, and $\sigma_{\mathrm{bond}} = 1$. The periodic simulation box extends from $-80\sigma$ to $80\sigma$ in each Cartesian direction, which is sufficiently large to treat the chain as isolated under the simulated conditions.

### 2.2 Langevin dynamics and tangential activity

The equations of motion are integrated with LAMMPS[35,36] . For bead $i$,

$$m\ddot{\mathrm{r}}_i =- \nabla_i U - \gamma\dot{\mathrm{r}}_i + \xi_i(t) + \mathrm{f}_i^{\mathrm{a}},$$

where $\gamma$ is the Langevin friction coefficient, $\xi_i$ is Gaussian thermal noise satisfying the fluctuation-dissipation relation, and $\mathrm{f}_i^{\mathrm{a}}$ is present only on active beads. The active force is tangent to the chain backbone,

$$\mathrm{f}_i^{\mathrm{a}} = f_a\,\hat{\mathrm{t}}_i, \qquad \hat{\mathrm{t}}_i = \frac{\mathrm{r}_{i+1} - \mathrm{r}_{i-1}}{|\mathrm{r}_{i+1} - \mathrm{r}_{i-1}|},$$

with the endpoint convention implemented by the custom LAMMPS fix. The signed Péclet number is

$$\mathrm{Pe} = \frac{f_a\sigma}{k_{\mathrm{B}}T}.$$

Positive and negative values denote propulsion toward increasing and decreasing bead index, respectively. For each point in Figures 1-8 and each architecture in Tables 1-2, the active group and signed Pe were changed to the corresponding bead interval(s) and polarity. All other model ingredients were kept consistent unless explicitly stated.

All simulations employ a timestep $\Delta t = 0.001$, a Langevin temperature $k_{\mathrm{B}}T = 1$, and a damping parameter $0.05$. The initial chain is relaxed with a limited-displacement integration and energy minimization, followed by an active Langevin relaxation of $10^7$ steps. The production trajectory is then sampled every 10,000 LAMMPS steps, corresponding to a saved-frame interval of 10 reduced time units. All production runs consist of $2 \times 10^9$ timesteps.

### 2.3 Activity architectures

A contiguous active block is specified by its inclusive bead-ID interval $[i_a, i_b]$ and by its signed propulsion direction. The number of active beads is $N_{\mathrm{active}} = i_b - i_a + 1$, and the active fraction is $p = \frac{N_{\mathrm{active}}}{N}$. The contour coordinate of the active-block center is defined as

$$s_{\mathrm{active}} = \frac{i_a + i_b}{2},$$

and its normalized position is $\frac{s_{\mathrm{active}}}{N}$.

The position scans in Figures 1–6 translate one contiguous active block along the contour while holding $p$ and the signed $Pe$ fixed. The quantity $\frac{s_{\mathrm{active}}}{N}$ is used for these single-block position scans; no single global block-center coordinate is assigned to the two-block architectures unless explicitly specified. The architecture scans in Figures 7–8 and Tables 1–2 instead identify each active block by its own bead interval and signed polarity.

For Table 1, $p = 0.4$ and $|\mathrm{Pe}| = 6$. The single block occupies beads 180-299 with positive polarity. The split architectures use beads 2-61 and 240-299, with polarity pairs $++$, $--$, $-+$, and $+-$ listed in increasing-bead-index order. For Table 2, $p = 0.3$ and $|\mathrm{Pe}| = 6$. The single block occupies beads 210–299 with positive polarity. The standard split geometry uses beads 2-46 and 180-224, while the terminal variant moves the second block to beads 255-299.

## 2.4 Knot detection and localization

Saved trajectories are analyzed with KymoKnot, which assigns a topological state to an open chain through a minimally interfering closure and locates the shortest knotted arc [37,38]. Knot localization in an open chain is intrinsically closure-dependent; the minimally interfering procedure is designed to reduce artificial entanglement introduced by the closure. Related definitions of polymer knot length motivate treating the localized arc as a fluctuating statistical object rather than a unique geometric invariant [39,40].

For each saved frame, the analysis records the knot type, the two boundaries of the shortest knotted arc, $s_1$ and $s_2$, its center,

$$s_K = \frac{s_1 + s_2}{2},$$

and its bead count,

$$L_K = s_2 - s_1 + 1.$$

KymoKnot outputs 0-based contour coordinates. For the plotted kymographs these coordinates were shifted by one so that the vertical coordinate is the 1-based LAMMPS bead ID. A knot episode is a strictly contiguous sequence of knotted saved frames. An episode is completed if both its first and last transitions are observed; episodes extending to the final saved frame are right-censored and are not included in the simple mean of completed lifetimes.

## 2.5 Statistical observables and two-state kinetic model

The steady-state knotting probability is

$$P_K = \frac{N_{\text{knotted frames}}}{N_{\text{all frames}}}.$$

The joint probabilities $P_{K,\text{act}}$ and $P_{K,\text{pas}}$ denote a knotted frame whose knot center lies in the active or passive region. Under the present two-region classification,

$$P_{K,\text{act}} + P_{K,\text{pas}} = P_K,$$

and the conditional passive residence is $P_{K,\text{pas}}/P_K$. The active fraction of the entire shortest knotted arc is

$$\phi_K^{\text{act}} = \frac{N_{\text{active beads in knot}}}{L_K}.$$

The conditional knot-formation rate, $\alpha_K^0$, is defined as the number of transitions from the unknotted state to the knotted state divided by the total accumulated time spent in the unknotted state,

$$\alpha_K^0 = \frac{N_{U \to K}}{T_U}.$$

To interpret the steady-state knotting probability, we use a minimal continuous-time two-state kinetic model (TSKM)[41], in which the polymer switches between an unknotted state $U$ and a knotted state $K$. The transition $U \to K$ occurs with the conditional formation rate $\alpha_K^0$. The reverse transition $K \to U$ is represented by the effective untying rate

$$\beta_K = \frac{1}{\langle \tau_K \rangle},$$

where $\langle \tau_K \rangle$ is the mean lifetime of completed knot episodes.

Let $P_K(t)$ denote the probability that the polymer is knotted at time $t$. The corresponding two-state rate equation is

$$\frac{dP_K}{dt} = (1 - P_K)\alpha_K^0 - \beta_K P_K.$$

At steady state, $\frac{dP_K}{dt} = 0$, so the probability flux from the unknotted state to the knotted state balances the reverse flux:

$$(1 - P_K^{(2)})\alpha_K^0 = \beta_K P_K^{(2)}.$$

Solving for the two-state prediction gives

$$P_K^{(2)} = \frac{\alpha_K^0}{\alpha_K^0 + \beta_K} = \frac{\alpha_K^0 \langle \tau_K \rangle}{1 + \alpha_K^0 \langle \tau_K \rangle}.$$

In the rare-event limit,

$$\alpha_K^0 \langle \tau_K \rangle \ll 1,$$

this expression reduces to

$$P_K^{(2)} \approx \alpha_K^0 \langle \tau_K \rangle.$$

This relation provides a kinetic baseline for interpreting the measured knotting statistics. Although its mathematical structure is that of a standard two-state stochastic process, its physical interpretation here is specific to polymer topology: $\alpha_K^0$ quantifies the efficiency of knot nucleation during unknotted intervals, whereas $\langle \tau_K \rangle$ quantifies knot survival or storage after formation. Consequently, changes in the measured steady-state probability $P_K$ can be separated into formation-controlled and lifetime-controlled contributions rather than being reported as a single occupancy alone.

Agreement between $P_K^{(2)}$ and the directly measured $P_K$ indicates that an effective two-state description is sufficient to rationalize the steady-state occupancy. Systematic deviations would instead point to dynamics not captured by the minimal model, such as simultaneous multiple knots, topology changes within a knotted episode, repeated capture before complete untying, activity-driven knot ejection, nonexponential lifetime statistics, or temporal correlations between formation and disappearance.

Bootstrap resampling is used to estimate standard errors of probabilities, means, and binned distributions [42]. For extremely rare cases, a zero or very small bootstrap error does not establish high precision; it may instead reflect one or a few detected frames.

# 3 Results and Discussion

## 3.1 Positive-propulsion terminal scans reveal source-to-reservoir transfer

Figure 1 resolves the kinetic origin of the position-dependent knotting response for $N = 300$, $p = 0.3$, and $Pe = +9$.

Figure 1(a) shows the directly measured steady-state knotting probability $P_K$ together with the two-state prediction $P_K^{(2)}$. Over most of the lower-$\frac{s_{active}}{N}$ range, knotting remains rare, although several local maxima occur. Once the active block enters the terminal regime, $P_K$ rises sharply and approaches order unity at the largest block-center positions. The close correspondence between $P_K$ and $P_K^{(2)}$ over both the low-probability and terminal regimes indicates that the large changes in occupancy are broadly accounted for by the measured formation rate and lifetime. Figure 1(b) shows the conditional passive-region residence probability $\frac{P_{K,pas}}{P_K}$. This quantity varies strongly with active-block position and reaches values close to unity for several terminal cases. Thus, even when activity is required to generate and transport the knot, the detected knot center spends a substantial, and often dominant, fraction of its knotted lifetime in the passive contour. The result excludes a picture in which the final knot remains permanently attached to the active block. Figure 1(c) displays the dimensionless conditional formation rate $\alpha_K^0 t$. In the lower-position range the formation rate is small and exhibits several local fluctuations. It then increases steeply as the active block approaches the terminal region, reaching its largest values near the upper end of the scanned range. The terminal rise in $P_K$ is therefore partly a nucleation effect: positioning the positively propelled block near the high-index free end increases the frequency with which the unknotted chain enters a knotted state. Figure 1(d) shows the mean lifetime $\langle\tau_K\rangle/t$ of completed knot episodes. The lifetime varies even more strongly than the formation rate and reaches values of order $10^4$–$10^5$ reduced time units in the terminal regime. Local lifetime peaks are also present at lower block positions, demonstrating that knot survival does not vary monotonically with nucleation efficiency. In particular, the largest terminal values of $P_K$ coincide with both an enhanced formation rate and a dramatic increase in lifetime. Consequently, terminal activity acts simultaneously on knot production and on the probability that a newly formed knot survives.

Taken together, Figures 1(a)–1(d) show that the terminal upsurge in knotting is not attributable to a single rate. The active block increases the supply of knotting events, whereas long residence in passive portions of the contour amplifies their contribution to the steady-state occupancy. Within the present position-scan data, this establishes the active-source/passive-storage separation that serves as a baseline for the polarity and architecture comparisons below. The corresponding division between active-region formation and passive-region retention is consistent with the terminal active-diblock mechanism reported in Ref. [31], while the subsequent sections test how that pathway changes when block position and signed polarity are independently controlled.

Figure 2 characterizes the geometry and composition of the knots underlying this kinetic response. Figure 2(a) shows the mean normalized knot-center position $\frac{\langle i_K\rangle}{N}$. If the knot center simply followed the translated active block, the data would lie on the diagonal reference line $\frac{\langle i_K\rangle}{N} = \frac{s_{active}}{N}$. Instead, substantial departures from this line occur, especially in the terminal regime. The mean knot center is therefore not rigidly advected with the active-block center; it can remain in, or be transported into, a different contour region. Figure 2(b) shows the mean normalized shortest-knotted-arc length $\frac{\langle L_K\rangle}{N}$. The strong, nonmonotonic variation between short and highly extended arcs demonstrates that the active-block position controls not only whether a knot exists but also whether the detected topology is tightly localized or delocalized over a large fraction of the contour. Figure 2(c) gives the mean active-bead fraction of the shortest knotted arc, $\langle \varphi_K^{act} \rangle$ . Values close to zero occur for several terminal configurations even when $P_K$ is large, showing that a high-knotting state need not be an active-dominated knot state. Conversely, larger values indicate configurations in which an extended knotted arc overlaps substantially with the active block. Comparison of Figures 1(b) and 2(c) is essential: the former classifies the location of a single knot-center coordinate, whereas the latter measures the composition of the complete shortest knotted arc.

Figure 3 resolves the full distributions hidden behind these averages. In the top row, corresponding to active beads 210–299, the knot-center distribution is broad and centered well below the active-block interval, while the knot-length distribution contains a strong short-length contribution together with a broad tail toward more extended knots. This combination is consistent with knots that are predominantly observed in the passive contour but retain a heterogeneous range of sizes. In the middle row, for active beads 206–295, the knot-center distribution is sharply concentrated near the terminal contour, and the knot-length distribution displays a pronounced localized population together with sparse extended events. This case represents a distinct subensemble rather than a smooth continuation of the top-row distribution. In the bottom row, for active beads 200–289, the knot-center distribution contains multiple maxima over the passive-side contour, while the knot-length distribution has a dominant short-knot population and a weaker secondary contribution at larger normalized knot length. These multiple peaks demonstrate coexistence between localized and more extended dynamical states. The full distributions therefore reveal broadening, multimodality, and long tails that cannot be reconstructed from the mean knot-center position and mean knot length alone.

## 3.2 Negative-propulsion scans separate nucleation from storage

Figure 4 gives the corresponding kinetic decomposition for $N = 300$, $p = 0.4$, and $Pe =- 5$, for which propulsion is directed toward decreasing bead index. Figure 4(a) shows that $P_K$ varies nonmonotonically with $\frac{s_{active}}{N}$. A first local enhancement occurs near $\frac{s_{active}}{N} \approx 0.70$, followed by a broader high-probability region around 0.74–0.77, after which $P_K$ decreases. The agreement between the measured and two-state values remains good, indicating that the position dependence can again be interpreted in terms of formation and survival. Figure 4(b) shows that $\frac{P_{K,pas}}{P_K}$ generally increases as the active block is displaced toward larger bead index, reaching approximately 0.45 at the end of the scan. The rise is not smooth, however, and local maxima occur at intermediate positions. The passive residence of the knot center is therefore geometry dependent and cannot be inferred solely from the distance of the active block from the free end. Figure 4(c) shows a broad maximum of $\alpha_K^0 t$ near $\frac{s_{active}}{N} \approx 0.75–0.76$. This identifies the block positions that most efficiently generate knotting events under negative propulsion. Figure 4(d) has a qualitatively different structure. The largest lifetime peak occurs closer to $\frac{s_{active}}{N} \approx 0.70$, with additional enhancement near smaller and intermediate block positions, followed by a gradual decline through the high-$\frac{s_{active}}{N}$ region. The formation-rate maximum and lifetime maximum therefore do not coincide. Similar values of $P_K$ can consequently arise either from relatively frequent knot formation or from relatively long knot survival. The mismatch between Figures 4(c) and 4(d) is direct evidence that nucleation and storage are independently tunable. It also explains why the steady-state knotting probability alone is insufficient for identifying the mechanism of a local maximum.

Figure 5 shows how the underlying knot geometry changes across the same scan. Figure 5(a) demonstrates that $\frac{\langle i_K \rangle}{N}$ does not simply translate with $\frac{s_{active}}{N}$. Although the mean center shifts overall toward larger bead index, its deviations from the active-block position show that knots can remain displaced from the driven region. Figure 5(b) shows substantial nonmonotonic changes in $\frac{\langle L_K \rangle}{N}$, with values ranging from moderately localized to strongly extended arcs. The lifetime enhancement in Figure 4(d) therefore does not correspond to a unique knot size. Figure 5(c) shows that $\langle \varphi_K^{act} \rangle$ also changes nonmonotonically. The knot may contain a large fraction of active beads even when its center is not located at the active-block center, confirming that center residence and arc composition describe different geometrical information.

Figure 6 resolves these changes at the distribution level. For the active block at 168–287, both the knot-center and knot-length populations are broad, indicating substantial heterogeneity. Moving the block to 157–276 redistributes probability between distinct contour and length intervals and generates secondary maxima. For 140–259, the distributions become more strongly structured, with localized peaks coexisting with long tails. The changes across the three rows therefore reflect redistribution between transient extended arcs and localized knots rather than a uniform translation of a single characteristic state.

### 3.3 Polarity reversal converts the same two-block architecture from a knot trap to a knot suppressor

Figure 7 and Table 1 isolate the effect of propulsion polarity at fixed chain length $N = 300$, active fraction $p = 0.4$, activity magnitude $|Pe| = 6$, and fixed active-block positions.

Figure 7(a) shows that the steady-state knotting probability is extraordinarily polarity sensitive. The Single + architecture has $P_K = 0.6875$. The Split ++ and Split -- cases have nearly equal intermediate probabilities, 0.1793 and 0.1786, respectively. Split -+, for which both terminal blocks propel toward their adjacent free ends, reaches $P_K = 0.9354$. In contrast, reversing both directions to Split +- reduces the probability to $3.897 \times 10^{-4}$. The two reversed-polarity architectures therefore differ by a factor of approximately $2.4 \times 10^{3}$, despite having identical active-bead positions and activity magnitude. Figure 7(b) shows that the conditional formation rate changes much less dramatically. Split -+ has $\alpha_K^0 t = 1.173 \times 10^{-4}$, whereas Split +- has $3.847 \times 10^{-5}$, a factor of only about 3.05. Moreover, the Single + case has an even larger formation rate than Split -+, although its steady-state $P_K$ is lower. The largest steady-state knotting probability is therefore not simply associated with the largest nucleation frequency. Figure 7(c) shows that the knot center resides predominantly in passive material for all of the well-sampled high- and intermediate-probability architectures. The passive-residence ratios are approximately 0.99 for Single +, Split ++, and Split --, and approximately 0.96 for Split -+. Even the rare Split +- detections give a predominantly passive-center ratio, although this value must be interpreted cautiously because of the extremely small number of knotted frames. Figure 7(d) identifies knot lifetime as the dominant polarity-sensitive quantity. The mean lifetime is approximately $1.22 \times 10^{5} t$ for Split -+, but only $10.13t$ for Split +-, a difference of approximately $1.2 \times 10^{4}$. Single +, Split ++, and Split -- have intermediate lifetimes. The polarity-induced switching of $P_K$ is therefore controlled primarily by survival rather than by formation frequency. Figure 7(e) adds the composition of the knotted arc. The stable high- and intermediate-probability cases have very small $\langle \varphi_K^{act} \rangle$, between approximately 0.013 and 0.040, demonstrating that their persistent knots are composed mainly of passive beads. By contrast, the fleeting Split +- events have $\langle \varphi_K^{act} \rangle \approx 0.404$. This high active fraction should not be interpreted as efficient active storage; it reflects short-lived detections that remain near active terminal regions and disappear before establishing a passive-reservoir state.

Taken together, Figures 7(a)–7(e) separate three stages. Polarity moderately changes how often knots form, strongly changes whether they survive, and determines whether the detected topology is transferred into passive material. The near symmetry of Split ++ and Split -- is also informative: each same-sign architecture aligns one terminal active block with a free end and misaligns the other, producing similar intermediate behavior. This suggests an approximate zero-one-two free-end-alignment rule for the symmetric terminal two-block geometry: zero, one, and two free-end-aligned active blocks correspond, respectively, to suppressed, intermediate, and high-knotting states. The Table 2 comparison nevertheless shows that this hierarchy is geometry-specific, because block placement provides an additional gate controlling knot formation and survival.

### 3.4 Terminal placement is a multiplicative gate on both nucleation and survival

Figure 8 and Table 2 examine $N = 300$, $p = 0.3$, and $|Pe| = 6$, and distinguish the effect of polarity from the effect of moving the second active block to a free end. Figure 8(a) shows that the Single + architecture has $P_K = 0.7257$, whereas the standard Split ++, Split --, Split -+, and Split +- geometries span probabilities from $5.060 \times 10^{-6}$ to 0.1222. The decisive comparison is between the interior Split -+ case, with $P_K = 5.060 \times 10^{-6}$, and Split -+ T, with $P_K = 0.6461$. Moving the positive second block from beads 180–224 to the terminal interval 255–299 therefore raises $P_K$ by approximately $1.28 \times 10^5$ without changing the polarity pair. Figure 8(b) shows that this terminal displacement strongly increases knot nucleation. The formation rate rises from $5.060 \times 10^{-7}$ for the interior Split -+ case to $6.697 \times 10^{-5}$ for Split -+ T, a factor of approximately 132. The terminal active block therefore opens an end-assisted formation channel that is absent in the corresponding interior geometry. Figure 8(c) shows that the high-probability states have different residence structures. Single + has an almost entirely passive-centered knot population, whereas Split -+ T has a passive-center ratio of approximately 0.67, leaving a substantial fraction of knot centers in active material. The rare interior Split -+ and Split +- events have ratios that are zero or poorly defined because their detected events are confined to one saved frame. These values are indicators of severe suppression rather than accurately sampled residence probabilities. Figure 8(d) shows that terminal placement also changes survival. The mean lifetime rises from one saved interval, $10t$, for interior Split -+ to approximately $2.73 \times 10^4 t$ for Split -+ T. Thus, the increase in $P_K$ is multiplicative: terminal placement enhances both the rate of entering the knotted state and the time spent in that state. Figure 8(e) reveals that Split -+ T has $\langle \varphi_K^{act} \rangle \approx 0.313$, substantially larger than the value for the Table 1 high-knotting Split -+ architecture. Single +, by contrast, has a very small active-arc fraction of approximately 0.014. High knotting therefore does not correspond to a unique spatial state. A high $P_K$ may arise from a predominantly passive stored knot, as in Single +, or from a compact knot population remaining closer to a terminal active block, as in Split -+ T.

These panel-by-panel comparisons demonstrate that free-end alignment is not by itself sufficient. The active block must also be placed so that the driven free end can generate threading configurations and the resulting knot can access a region in which it survives. Polarity and block position therefore act jointly rather than additively.

### 3.5 Individual trajectories reveal formation, transport, tightening, and escape

Figures 9–11 provide the single-episode dynamics corresponding to the statistical contrasts in Figure 7 and Table 1.

Figure 9(a) shows a complete topology-pure trefoil episode for the Single + architecture. The knot is first detected near the terminal active block, with the knotted arc spanning beads 252–300, and subsequently travels across almost the full contour before disappearing near beads 1–23. This trajectory demonstrates long-range contour transport rather than stationary localization at the formation site. Figure 9(b) shows an initial trefoil segment in which the knot center moves from bead 217.5 in the active region to bead 100.5 in the passive region, while the shortest knotted arc contracts from 166 to 24 beads. Migration across the active–passive interface is therefore accompanied by pronounced tightening. Figure 9(c) shows a long topology-homogeneous segment with the same net direction of motion: the knot center moves from approximately bead 211.5 to bead 41 while the arc length decreases from 158 to 61 beads. The repeated transport direction in panels (a)–(c) indicates that the conveyor-like motion is not restricted to a single episode or knot type.

Figure 10 concerns the high-knotting Split -+ architecture. Panel (a) is a one-frame topology detection near the terminal active block and therefore represents a failed or unresolved precursor rather than sustained transport. Panel (b) is a second short episode lasting four saved frames immediately before the dominant long-lived state. The occurrence of these brief events suggests repeated threading attempts near the driven terminal region. Figure 10(c) shows the initial topology-pure portion of the dominant episode. The knot forms in the terminal active block, its center migrates toward decreasing bead index, crosses the interface at bead coordinate 239.5, and enters the central passive region while the shortest knotted arc contracts. The sequence of short precursors followed by successful long-lived capture is consistent with a stochastic nucleation process coupled to passive-region stabilization.

Figure 11 shows the polarity-reversed Split +- architecture. Panel (a) contains the longest detected episode, but it persists for only two saved frames. During those two observations the knotted arc expands from 151 to 195 beads rather than tightening. Panel (b) is a one-frame trefoil detection near the lower-index active block, and panel (c) is a one-frame detection near the terminal active block. Neither event develops into persistent contour transport.

The complete Split +- trajectory contains 78 knotted frames distributed among 77 episodes: 76 episodes last one saved frame and the longest lasts two. The contrast with Figures 9 and 10 is therefore qualitative, not merely quantitative. Free-end-aligned polarities permit repeated formation followed by migration and localization, whereas the reversed polarities generate only fleeting topology signatures. These trajectories directly establish that active fraction, block number, and block positions alone cannot explain the knotting state.

### 3.6 Polarity-programmed source–conveyor–reservoir–sink mechanism

Figure 12 integrates the kinetic, spatial, architectural, and episode-resolved evidence into a polarity-programmed source–conveyor–reservoir–sink mechanism. A free-end-aligned terminal block acts as a topological source, tangentially driven contour conveys a newly formed knot away from its formation region, passive contour provides a long-lived reservoir, and reversed or unfavorably positioned blocks interrupt the pathway as effective barriers or sinks. The terminal-diblock limit is consistent with the formation–migration–retention phenomenology of Ref. [31], while the present results establish that independently assigned block positions and signed polarities can open, redirect, or close the complete pathway.

Figure 12(a) represents the high-knotting, free-end-aligned configuration. Outward propulsion makes the adjacent free ends active leading tips and increases access to end-backfolding and end–loop threading configurations. A successfully formed knot is then transported away from the terminal active block, crosses an active–passive interface, and enters passive contour, where its shortest knotted arc can localize and its lifetime increases. Figure 12(b) represents the polarity-reversed, free-end-misaligned configuration. Propulsion is directed away from the threadable free ends and toward the interior interfaces. In the present trajectories this arrangement produces only short-lived topology detections and does not establish sustained transfer into a passive storage region. The reversed architecture therefore does more than reduce the knotting probability: it closes the long-lived transport-and-retention pathway that is open in Figure 12(a). Interior placement can interrupt the same pathway even when the polarity pair is unchanged, as shown by the terminal-placement comparison in Figure 8 and Table 2.

The results can be organized into four mechanistic elements.

1. Terminal active blocks act as topology sources only when polarity is matched to the free end. A linear chain becomes knotted through passage of a free end through a loop[3,5,7]. When a terminal active block points toward its adjacent free end, that end is also the active leading tip. Persistent tangential forcing then promotes backfolding and repeated end-loop encounters. In the terminal split geometry, $-+$ aligns both blocks with their nearest free ends, $++$ and $--$ align one, and $+-$ aligns neither. The sequence high-intermediate-low in Table 1 follows this ordering.

2. Knot motion is opposite to simple attachment to the active block. The knot center frequently leaves the region in which the knot first appears. Kymographs show directed contour transport from a terminal active domain into the passive interior. This behavior is consistent with the knot acting as a slip-link-like topological object through which actively driven contour is fed. The knot therefore drifts relative to bead labels even though the polymer as a whole is not pulled by an external force[11,12,31].

3. Passive regions are topology reservoirs. High-$P_K$ states have large passive-center residence and small active-arc fractions. Once a knot enters a passive domain, reduced active contour throughput allows the arc to tighten and makes return to a free end slow. This storage step increases $\langle\tau_K\rangle$ by orders of magnitude and dominates the polarity contrast in Table 1[31].

4. Misaligned or interior active blocks can act as topology sinks or barriers. When propulsion points away from a nearby free end, the active leading edge lies at an interface rather than at a threadable chain end. Formation is reduced. If an interior block drives contour in a direction that returns a knot toward its formation end or prevents entry into a passive reservoir, survival is also reduced. The Table 2 terminal-placement experiment shows that this barrier/sink role is geometry-dependent, explaining why free-end matching is necessary but not sufficient for arbitrary block arrangements.

The contrast between Figures 12(a) and 12(b) supports describing the spatial activity pattern as a polarity-controlled topological gate and, for the terminal split geometry, as diode-like with respect to polarity reversal. This terminology is operational: one orientation supports knot formation, directed contour transfer, and long-lived retention, whereas the reversed orientation suppresses sustained transport and storage. It does not imply an electrical-diode-like microscopic force law, nor does it identify a specific local interface force without further bead-resolved dynamical measurements.

### 3.7 Scope and limitations of the mechanistic interpretation

The evidence establishes strong correlations among architecture, signed polarity, formation rate, lifetime, residence, and single-episode transport. It does not yet directly measure a local mobility discontinuity, contour current, bond tension, or stress at the active-passive interface. The source–conveyor–reservoir–sink framework should therefore be further tested by bead-resolved mobility, signed tangential displacement, bond-extension or bonded-force proxies, and event-aligned curvature/contact observables. Such measurements can distinguish whether the interface acts primarily as a mobility boundary, a force-transmission boundary, a geometric folding zone, or a kinetic trapping region.

Several additional limitations should be retained. First, most results are for $N = 300$; finite-size scaling is required to determine whether the polarity contrast persists or sharpens with chain length[1,2,3,4,5,40]. Second, hydrodynamic interactions are neglected[14,16,26]. Third, KymoKnot localization is robust but not closure-independent for open chains[37,38]. Fourth, rare-event cases with one-frame episodes do not support stable conditional distributions.

## 4 Conclusions

Spatially patterned tangential activity controls polymer knotting through the coupled effects of block position and signed polarity, which regulate knot formation, contour transport, and survival. The single-block position scans show that moving an active domain changes not only the steady-state knotting probability $P_K$ but also the conditional formation rate, mean lifetime, passive-region residence, knot-center location, and knotted-arc composition. The close agreement between the directly measured $P_K$ and the two-state kinetic-model prediction $P_K^{(2)}$ shows that the large variation in $P_K$ can be rationalized primarily by the competition between formation and untying kinetics.

The principal result is that signed polarity acts as an independent topological control variable. For $N = 300$, $p = 0.4$, and $|Pe| = 6$, two split architectures with identical active-bead positions but reversed polarities differ in $P_K$ by approximately $2.4 \times 10^3$. Their conditional formation rates differ by only a factor of 3.05, whereas their mean knot lifetimes differ by approximately $1.2 \times 10^4$. The polarity switch is therefore dominated by the opening or closure of a long-lived transport-and-retention pathway rather than by knot nucleation alone. Geometry provides an additional gate: for $N = 300$, $p = 0.3$, and $|Pe| = 6$, moving the positive second block from an interior interval to the terminal interval raises $P_K$ by approximately $1.28 \times 10^5$ through simultaneous enhancement of formation and survival.

Beyond this binary switch, the symmetric terminal two-block architectures reveal an approximate zero-one-two free-end-alignment rule. Split +- aligns neither terminal active block with its adjacent free end and produces the suppressed-knotting state ($P_K$ = $3.897 \times 10^{-4}$); Split ++ and Split -- each align one terminal block and yield nearly identical intermediate knotting probabilities ($P_K$ = 0.1793 and 0.1786, respectively); and Split -+ aligns both terminal blocks and yields the high-knotting state ($P_K$ = 0.9354). Thus, within this symmetric terminal geometry, the number of free-end-aligned active blocks provides a compact predictor of the knotting hierarchy. This rule is geometry-specific rather than universal: the terminal-placement comparison shows that block position supplies an additional gate controlling both knot formation and survival.

Together, the polarity switch, the zero-one-two free-end-alignment rule, and the terminal-placement comparison establish a programmable topological gating principle and support a polarity-programmed source–conveyor–reservoir–sink mechanism. In the studied terminal geometries, a free-end-aligned active block promotes end-assisted threading and acts as a topological source; tangentially driven contour conveys newly formed knots away from the formation region; passive contour provides a dynamically quieter reservoir in which knots can localize and persist; and reversed or unfavorably placed active blocks interrupt the pathway as effective barriers or sinks. Independently assigning active-block positions and polarities therefore selects whether the chain predominantly nucleates, transports, stores, or rapidly loses knots.

The proposed mechanism is supported by the kinetic decomposition, passive-residence and arc-composition statistics, the zero-one-two hierarchy, the terminal-placement control, and episode-resolved kymographs. Bead-resolved mobility and bonded-force proxies, event-aligned precursor analysis, finite-size tests, and broader scans of activity and block geometry will further resolve the microscopic interface dynamics and test the generality of the gating behavior. Overall, spatial activity patterning provides a robust strategy for programming whether a short active polymer forms, transports, stores, or suppresses knots.

Acknowledgement

This work is supported in part by the High Performance Computing Center of Central South University.

Figure 1. Position-dependent kinetic decomposition for a chain with $N = 300$, $p = 0.3$, and $Pe =+ 9$. The active block is translated along the contour, and $\frac{s_{\mathrm{active}}}{N}$ denotes its normalized center position. (a) Directly measured steady-state knotting probability $P_K$ (Sim) and the two-state kinetic-model prediction $P_K^{(2)}$ (Sim-TSKM). (b) Conditional passive-region residence probability $\frac{P_{K,\mathrm{pas}}}{P_K}$ of the knot center. (c) Dimensionless conditional knot-formation rate $\alpha_K^0 t$. (d) Mean lifetime $\frac{\langle \tau_K \rangle}{t}$ of completed knot episodes. Main panels magnify the terminal-block regime, whereas insets show the lower-position range. Error bars are bootstrap standard errors where available.

Figure 2. Position-dependent spatial localization and knot-arc composition for $N = 300$, $p = 0.3$, and $Pe =+ 9$. The active block is translated along the contour, with horizontal coordinate $\frac{s_{\mathrm{active}}}{N}$. (a) Mean normalized knot-center position $\frac{\langle i_K \rangle}{N}$; the green diagonal is the reference relation $\frac{\langle i_K \rangle}{N} = \frac{s_{\mathrm{active}}}{N}$, expected if the knot center follows the active-block center. (b) Mean normalized shortest-knotted-arc length $\frac{\langle L_K \rangle}{N}$. (c) Mean active-bead fraction $\langle \varphi_K^{\mathrm{act}} \rangle$ of the shortest knotted arc. Main panels magnify the terminal-block regime, whereas insets show the lower-position range. All averages are conditional on knotted frames. Error bars are bootstrap standard errors.

Figure 3. Representative binned knot-position and knot-length distributions for terminal-region single-block activity at $N = 300$, $p = 0.3$, and $Pe =+ 9$. The left and right columns show $\mathrm{P}(\frac{i_K}{N})$ and $\mathrm{P}(\frac{L_K}{N})$, respectively. From top to bottom, the rows correspond to positive active blocks occupying beads 210-299, 206-295, and 200-289. Ordinates are bin probabilities rather than probability densities, and the probabilities in each distribution sum to unity. Error bars are bootstrap standard errors.

Figure 4. Position-dependent kinetic decomposition for $N = 300$, $p = 0.4$, and $Pe =- 5$, with propulsion toward decreasing bead index. The horizontal coordinate is the normalized active-block center $\frac{s_{\mathrm{active}}}{N}$. Panels show (a) Directly measured steady-state

knotting probability $P_K$ (Sim) and the two-state kinetic-model prediction $P_K^{(2)}$ (Sim-TSKM), (b) $\frac{P_{K,\mathrm{pas}}}{P_K}$, (c) $\alpha_K^0 t$, and (d) $\frac{\langle \tau_K \rangle}{t}$. The formation-rate and lifetime extrema occur at different block positions, showing that nucleation and survival vary independently. Error bars are bootstrap standard errors where available.

Figure 5. Position-dependent spatial localization and knot-arc composition for $N = 300$, $p = 0.4$, and $Pe =- 5$. Panels show (a) the mean normalized knot-center position $\frac{\langle i_K \rangle}{N}$, (b) the mean normalized shortest-knotted-arc length $\frac{\langle L_K \rangle}{N}$, and (c) the mean active-bead fraction $\langle \varphi_K^{\mathrm{act}} \rangle$ of the shortest knotted arc. The green diagonal in panel (a) is the reference relation $\frac{\langle i_K \rangle}{N} = \frac{S_{\mathrm{active}}}{N}$. All averages are conditional on knotted frames. Error bars are bootstrap standard errors.

Figure 6. Representative binned knot-position and knot-length distributions for $N = 300$, $p = 0.4$, and $Pe =- 5$. The left and right columns show $\mathrm{P}(\frac{i_K}{N})$ and $\mathrm{P}(\frac{L_K}{N})$, respectively. From top to bottom, the rows correspond to negative active blocks occupying beads 168-287, 157-276, and 140-259. Each distribution is normalized separately so that its bin probabilities sum to unity. The changing widths, multiple maxima, and long tails resolve localization changes that are hidden by the averages in Figure 5. Error bars are bootstrap standard errors.

Figure 7. Architecture- and polarity-dependent knotting for $N = 300$, $p = 0.4$, and $|Pe| = 6$. The Single + architecture contains one positive active block on beads 180-299. The split architectures contain active blocks on beads 2-61 and 240-299; polarity pairs are listed in increasing-bead-index order, with + and - denoting propulsion toward increasing and decreasing bead index, respectively. Panels show (a) $P_K$, (b) $\alpha_K^0 t$, (c) $\frac{P_{K,\mathrm{pas}}}{P_K}$, (d) $\frac{\langle \tau_K \rangle}{t}$, and (e) $\langle \varphi_K^{\mathrm{act}} \rangle$. At fixed active fraction, activity magnitude, and block positions, reversing Split -+ to Split +- changes $P_K$ by approximately $2.4 \times 10^3$. This contrast is dominated by an approximately $1.2 \times 10^4$-fold difference in mean lifetime, whereas the formation rates differ by only a factor of 3.05. Error bars are bootstrap standard errors.

Figure 8. Architecture-, polarity-, and block-position-dependent knotting for $N = 300$, $p = 0.3$, and $|Pe| = 6$. The Single + architecture contains one positive active block on beads 210-299. The standard split geometry contains active blocks on beads 2-46 and 180-224, whereas Split -+ T moves the positive second block to the terminal interval 255-299 while retaining the first block and the -+ polarity pair. Polarity pairs are listed in increasing-bead-index order, with + and - denoting propulsion toward increasing and decreasing bead index, respectively. Panels show (a) $P_K$, (b) $\alpha_K^0 t$, (c) $\frac{P_{K,\text{pas}}}{P_K}$, (d) $\frac{\langle \tau_K \rangle}{t}$, and (e) $\langle \varphi_K^{\text{act}} \rangle$. The controlled comparison between interior Split -+ and terminal Split -+ T isolates the effect of terminal placement, which increases $P_K$ by approximately $1.28 \times 10^5$ through simultaneous enhancement of knot formation and survival. Error bars are bootstrap standard errors.

Figure 9. Representative knot-episode kymographs for the Table 1 Single + architecture, with positive propulsion on beads 180-299. The blue and orange curves are the two boundaries $s_1(\mathrm{t})$ and $s_2(\mathrm{t})$ of the shortest knotted arc, the green dashed curve is its center $s_K(\mathrm{t})$, and the filled region is the instantaneous shortest knotted arc. Gray shading marks the active block, the horizontal dashed line marks the active-passive boundary, and time is measured from the beginning of each displayed segment. (a) A complete topology-pure trefoil episode in which the knot migrates from the terminal active block across nearly the full contour. (b) An initial trefoil segment in which active-to-passive transport is accompanied by contraction of the arc from 166 to 24 beads. (c) A long topology-homogeneous 5_2 segment showing the same net transport direction and contraction of the arc from 158 to 61 beads.

Figure 10. Representative knot-episode kymographs for the high-knotting Table 1 Split -+ architecture, with active blocks on beads 2-61 (-) and 240-299 (+). The curves show $s_1(\mathrm{t})$, $s_2(\mathrm{t})$, and $s_K(\mathrm{t})$; shaded terminal bands mark the active blocks, the unshaded central interval is passive, and horizontal dotted lines mark the interfaces at bead coordinates 61.5 and 239.5. Time is measured relative to the beginning of each episode. (a) A one-frame trefoil detection near the upper terminal active block. (b) A four-frame precursor immediately preceding the dominant long-lived state. (c) The initial topology-pure trefoil portion of the dominant episode, showing formation in the upper active block, migration across the interface at 239.5 into the passive region, and net tightening of the shortest knotted arc.

Figure 11. Representative event-centered kymographs for the polarity-suppressed Table 1 Split +- architecture, with active blocks on beads 2-61 (+) and 240-299 (-). The symbols show the shortest-arc boundaries $s_1(\mathrm{t})$ and $s_2(\mathrm{t})$ and the knot center $s_K(\mathrm{t})$; filled regions denote the shortest knotted arc, shaded terminal bands mark the active blocks, and horizontal dashed lines mark the two active-passive interfaces. (a) The longest detected episode persists for only two saved frames and expands from 151 to 195 beads. (b,c) One-frame detections near the lower-index and upper terminal active blocks; the adjacent saved frames are unknotted. Across the full trajectory, 78 knotted frames form 77 episodes, of which 76 last for one saved frame, demonstrating transient topology detection rather than sustained transport or passive-region storage.

Figure 12. Proposed polarity-programmed source-conveyor-reservoir-sink mechanism. Orange and blue segments denote active and passive contour, respectively; black arrows indicate local propulsion direction, and red symbols indicate the inferred knot-formation, transport, retention, and escape processes. (a) Outward, free-end-aligned terminal polarities create active leading tips that favor end-assisted threading, convey a newly formed knot into passive contour, and support tightening and long-lived storage; this orientation corresponds to an open topological gate. (b) Reversed, free-end-misaligned polarities direct propulsion away from the threadable free ends, interrupt sustained transfer into the passive reservoir, and produce rapid disappearance or end escape; this orientation corresponds to a closed gate or an effective topological sink. The schematic summarizes the operational interpretation supported by Figures 7-11 and is not a direct measurement of local force, stress, mobility, or contour current.

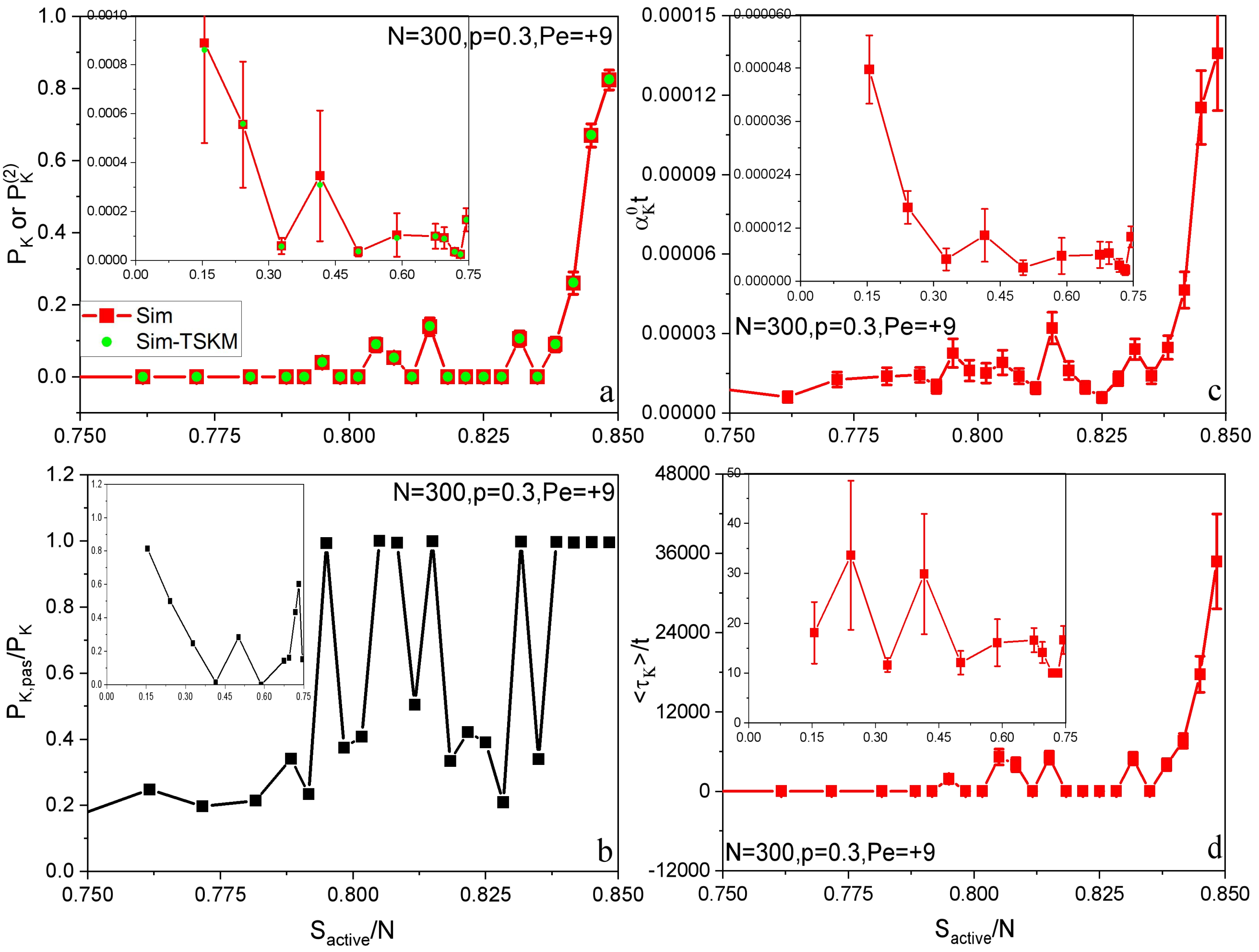

N=300,p=0.3,Pe=+9
$P_K$ or $P_K^{(2)}$
Sim
Sim-TSKM
a
$\alpha_K^0 t$
c
$P_{K,pas}/P_K$
b
$<\tau_K>/t$
d
$S_{active}/N$

Figure 1

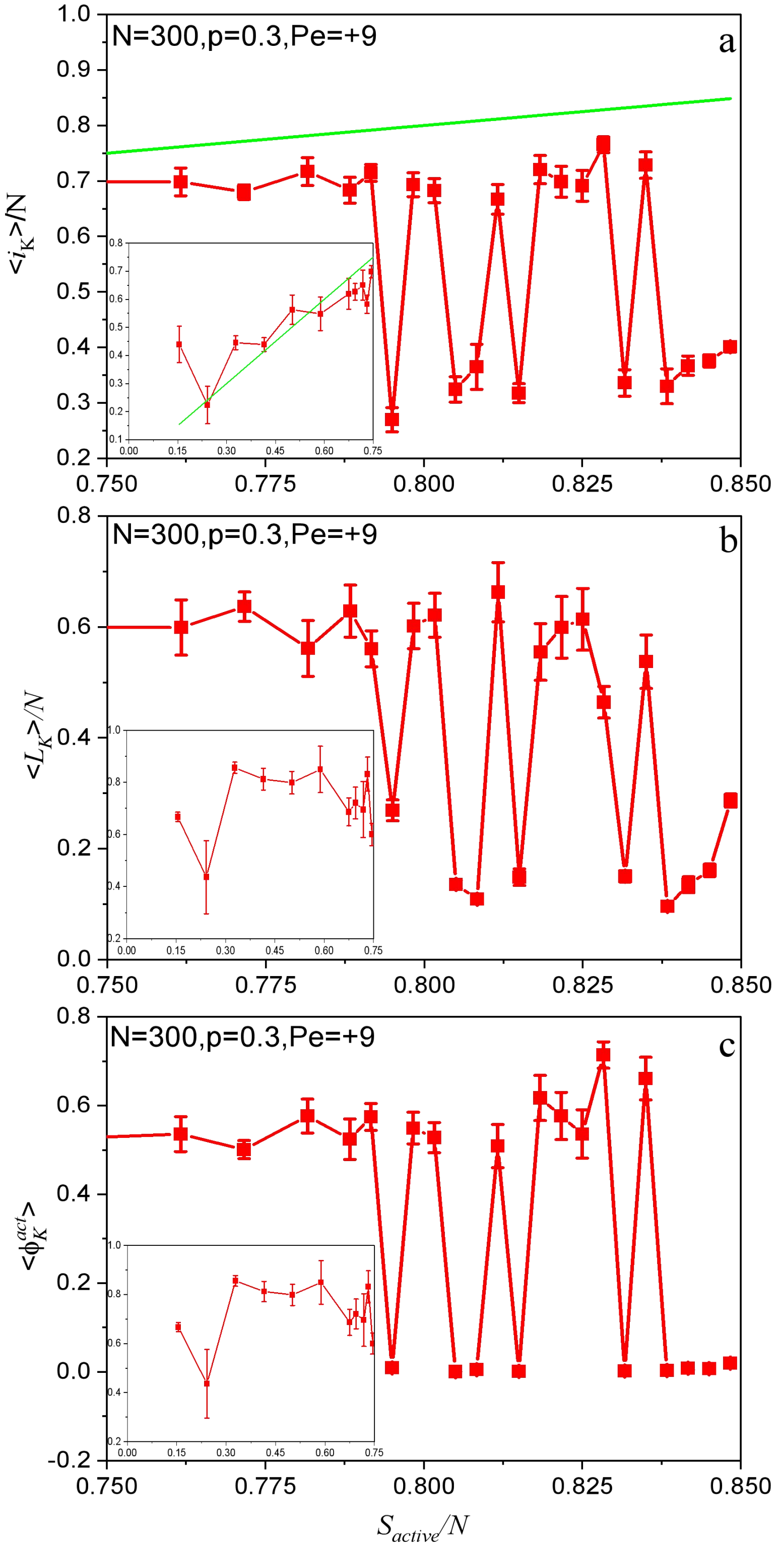
N=300,p=0.3,Pe=+9
a
<i_K>/N
N=300,p=0.3,Pe=+9
b
<L_K>/N
N=300,p=0.3,Pe=+9
c
<φ_K^act>
S_active/N

Figure 2

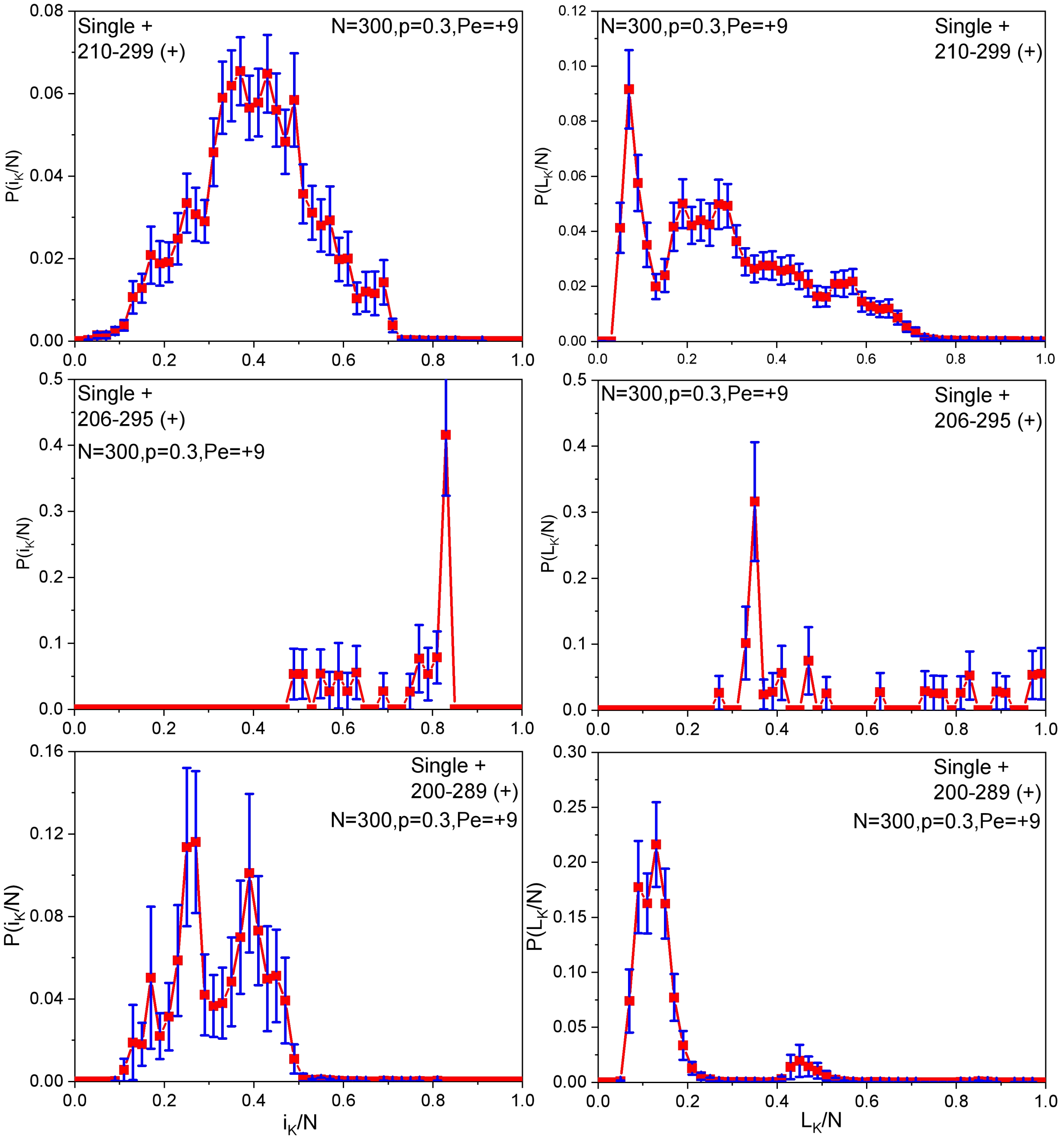
Single +
210-299 (+)
N=300,p=0.3,Pe=+9
P(iK/N)
Single +
210-299 (+)
N=300,p=0.3,Pe=+9
P(LK/N)
Single +
206-295 (+)
N=300,p=0.3,Pe=+9
P(iK/N)
N=300,p=0.3,Pe=+9
Single +
206-295 (+)
P(LK/N)
Single +
200-289 (+)
N=300,p=0.3,Pe=+9
P(iK/N)
iK/N
Single +
200-289 (+)
N=300,p=0.3,Pe=+9
P(LK/N)
LK/N

Figure 3

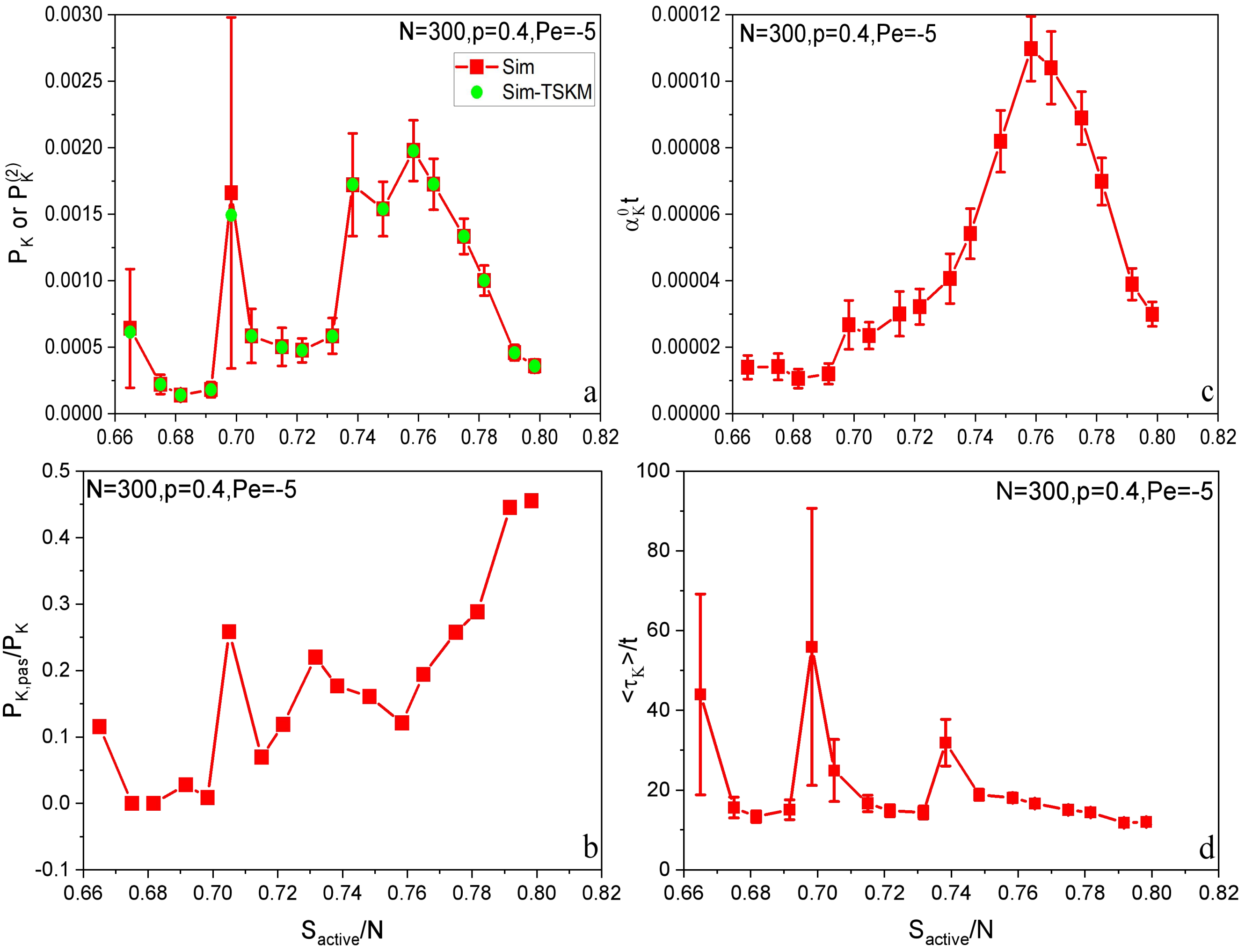

N=300,p=0.4,Pe=-5
Sim
Sim-TSKM
P_K or P_K^(2)
a
α_K^0 t
c
P_K,pas/P_K
b
<τ_K>/t
d
S_active/N
S_active/N

Figure 4

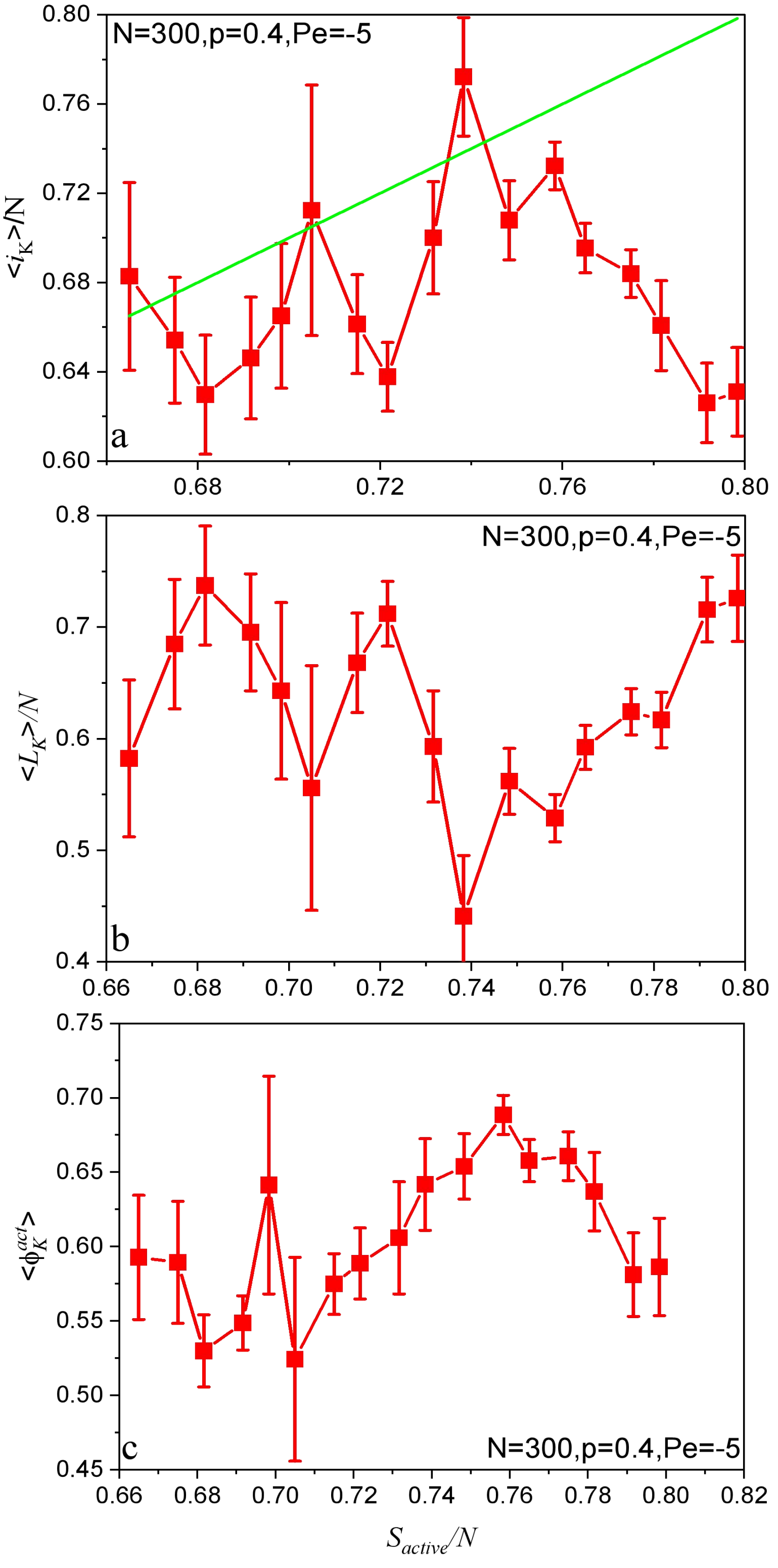
N=300,p=0.4,Pe=-5
<$i_K$>/N
a
N=300,p=0.4,Pe=-5
<$L_K$>/N
b
N=300,p=0.4,Pe=-5
<$\phi_K^{act}$>
c
$S_{active}/N$

Figure 5

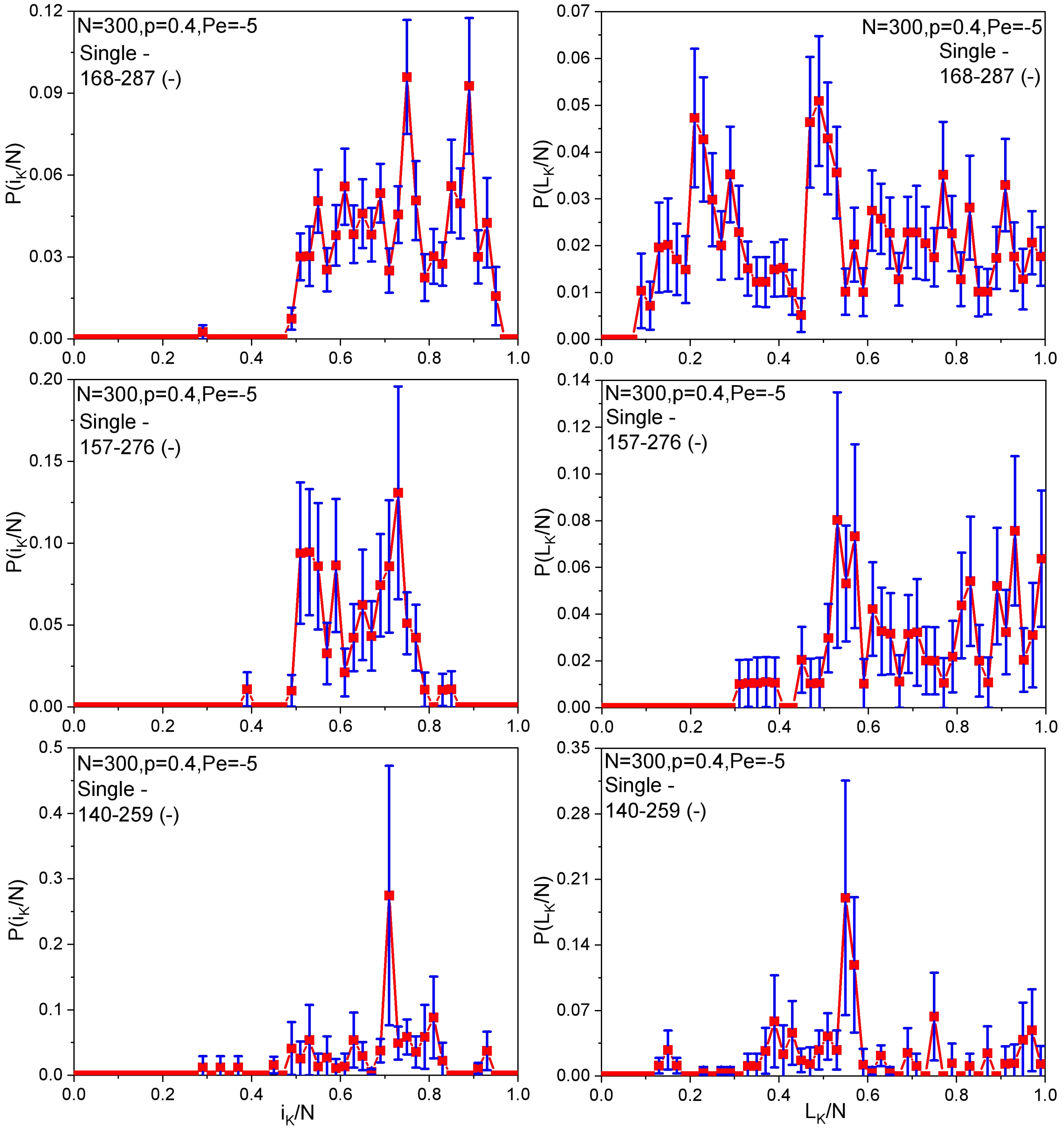

N=300,p=0.4,Pe=-5
Single -
168-287 (-)
P(i_K/N)
N=300,p=0.4,Pe=-5
Single -
168-287 (-)
P(L_K/N)
N=300,p=0.4,Pe=-5
Single -
157-276 (-)
P(i_K/N)
N=300,p=0.4,Pe=-5
Single -
157-276 (-)
P(L_K/N)
N=300,p=0.4,Pe=-5
Single -
140-259 (-)
P(i_K/N)
i_K/N
N=300,p=0.4,Pe=-5
Single -
140-259 (-)
P(L_K/N)
L_K/N

Figure 6

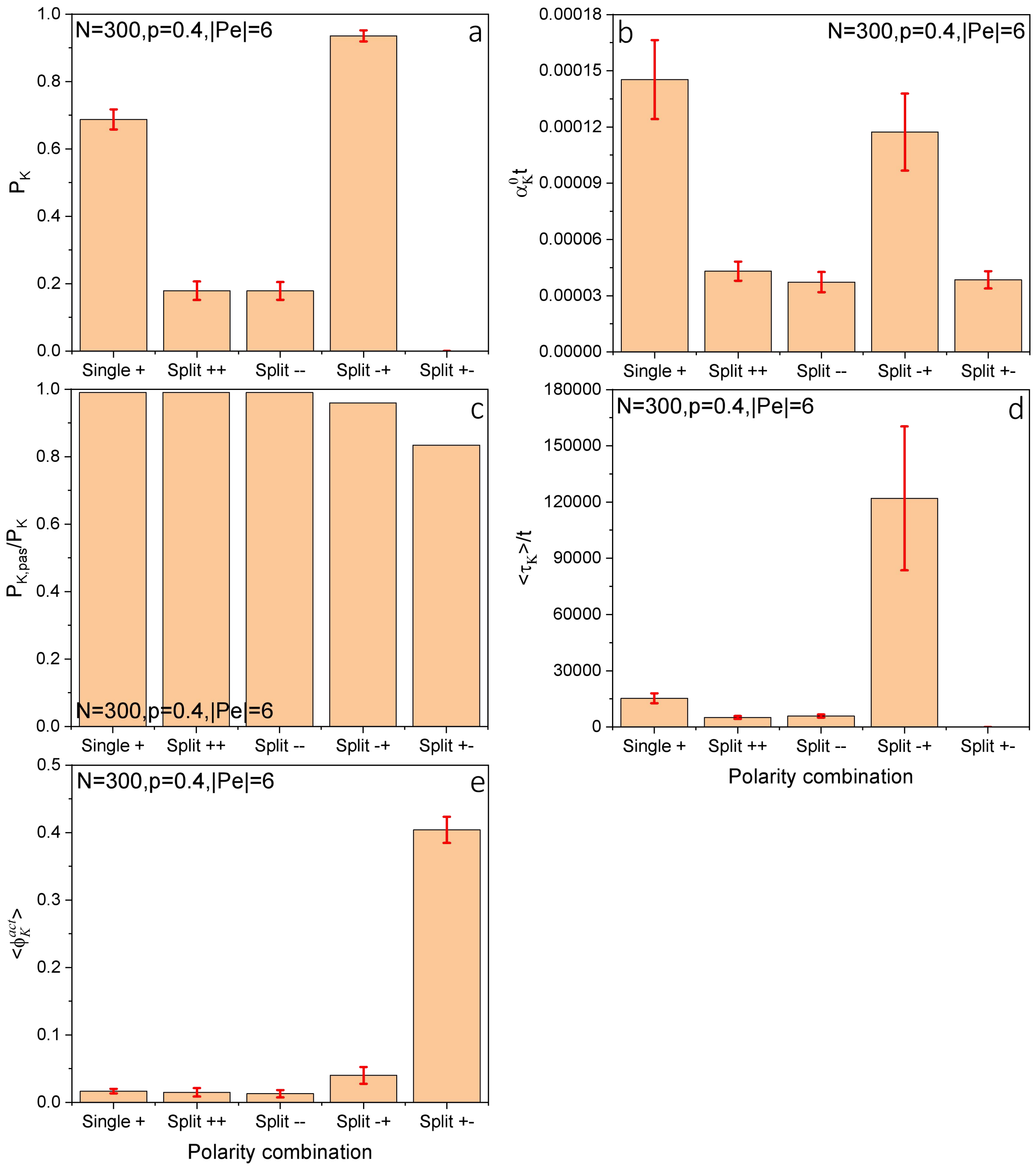
N=300,p=0.4,|Pe|=6
a
b
c
d
e
$P_K$
$\alpha_K^0 t$
$P_{K,pas}/P_K$
$<\tau_K>/t$
$<\phi_K^{act}>$
Single +
Split ++
Split --
Split -+
Split +-
Polarity combination

Figure 7

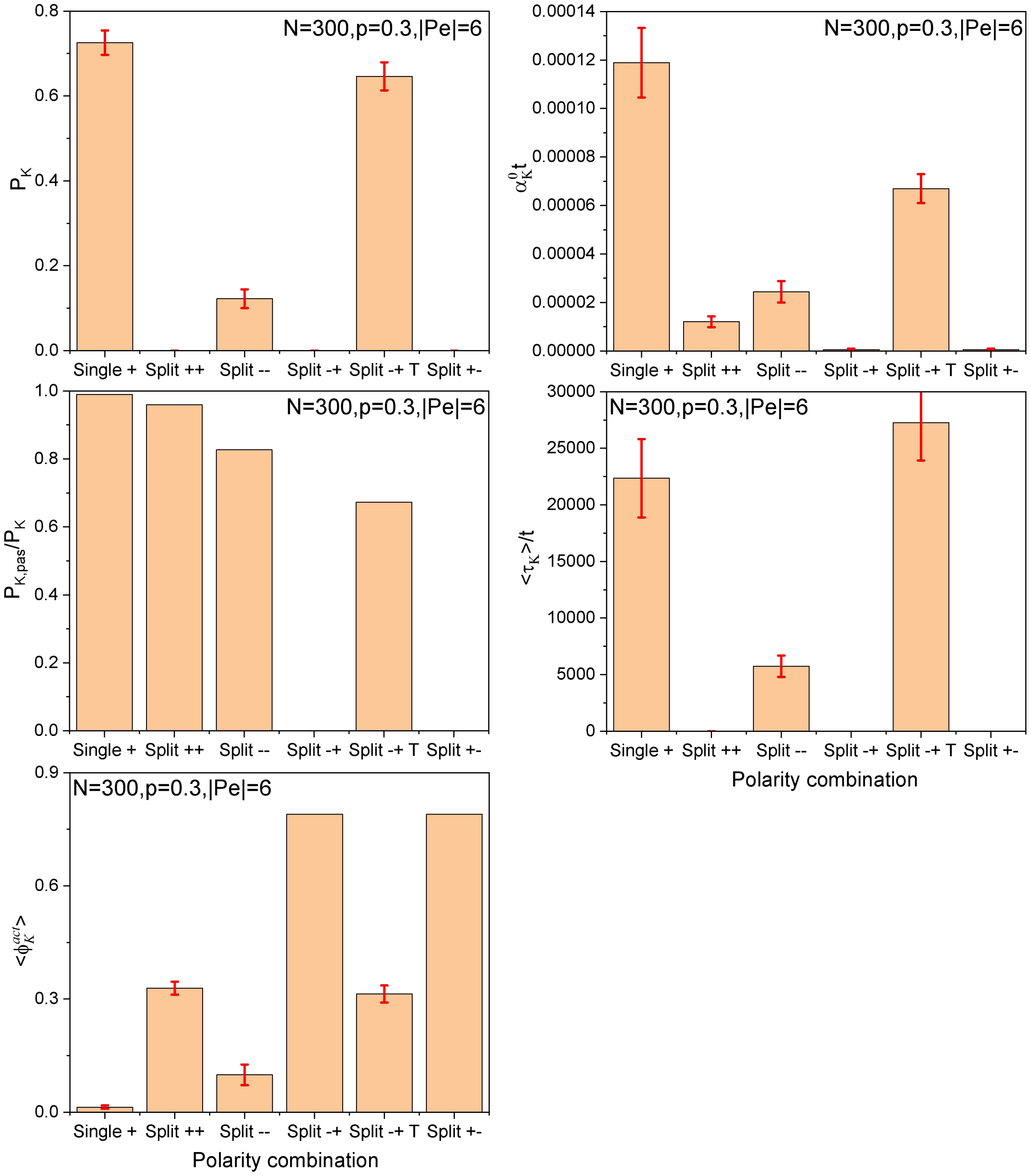
N=300,p=0.3,|Pe|=6
$P_K$
$\alpha_K^0 t$
$P_{K,pas}/P_K$
$<\tau_K>/t$
$<\phi_K^{act}>$
Single +
Split ++
Split --
Split -+
Split -+ T
Split +-
Polarity combination

Figure 8

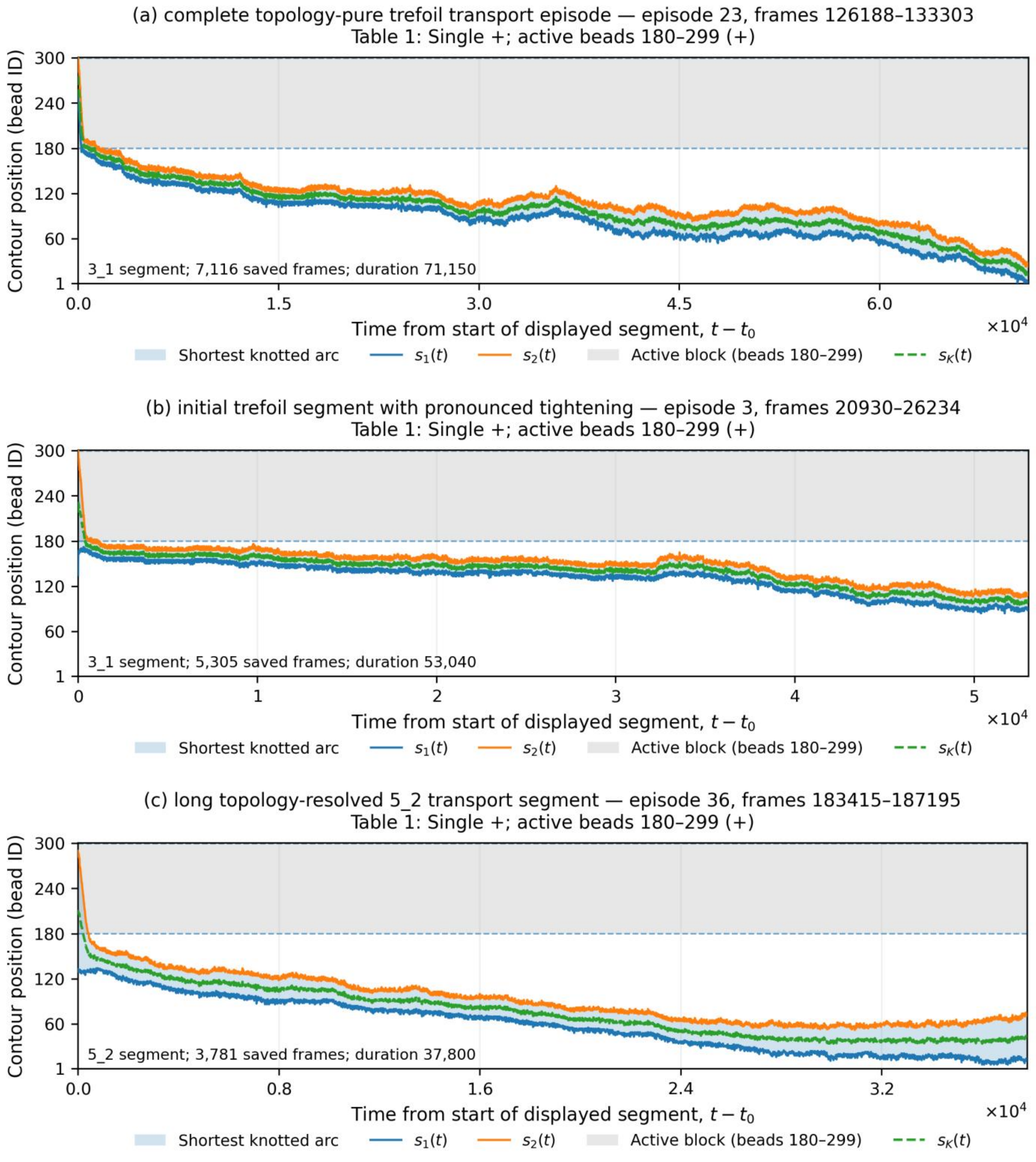

(a) complete topology-pure trefoil transport episode — episode 23, frames 126188–133303
Table 1: Single +; active beads 180–299 (+)
Contour position (bead ID)
3_1 segment; 7,116 saved frames; duration 71,150
Time from start of displayed segment, t − t0
×10^4
Shortest knotted arc
s1(t)
s2(t)
Active block (beads 180–299)
sK(t)
(b) initial trefoil segment with pronounced tightening — episode 3, frames 20930–26234
Table 1: Single +; active beads 180–299 (+)
Contour position (bead ID)
3_1 segment; 5,305 saved frames; duration 53,040
Time from start of displayed segment, t − t0
×10^4
Shortest knotted arc
s1(t)
s2(t)
Active block (beads 180–299)
sK(t)
(c) long topology-resolved 5_2 transport segment — episode 36, frames 183415–187195
Table 1: Single +; active beads 180–299 (+)
Contour position (bead ID)
5_2 segment; 3,781 saved frames; duration 37,800
Time from start of displayed segment, t − t0
×10^4
Shortest knotted arc
s1(t)
s2(t)
Active block (beads 180–299)
sK(t)

Figure 9

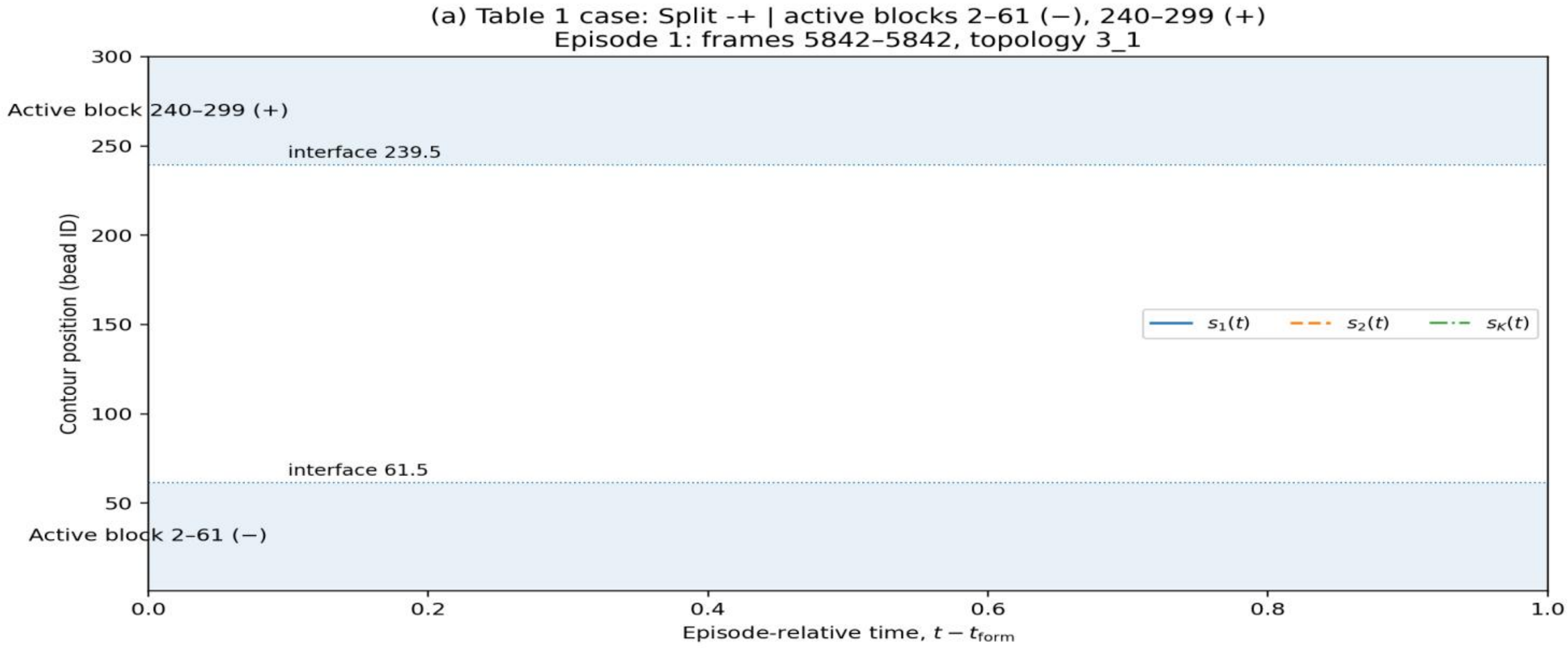

Short transient knot episode near the terminal active block.

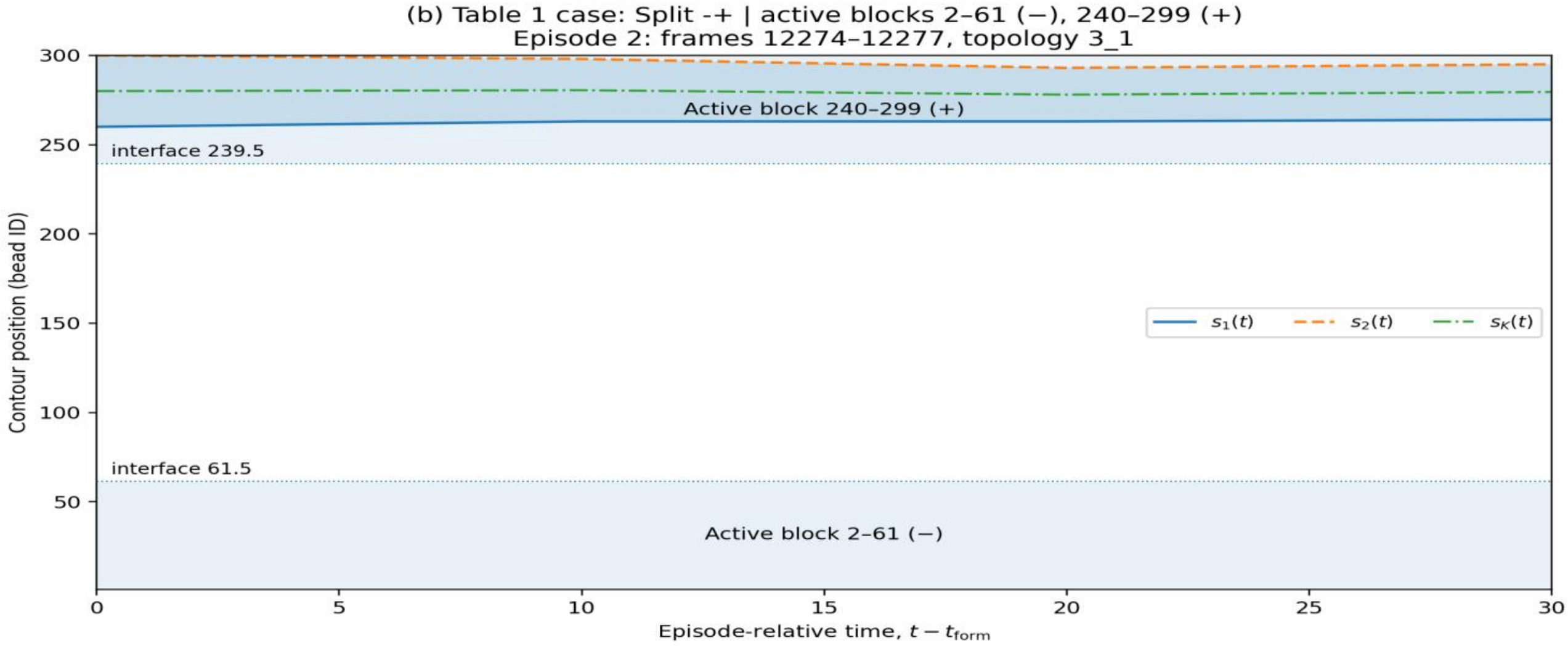

Another short transient knot episode preceding the long-lived knotted state.

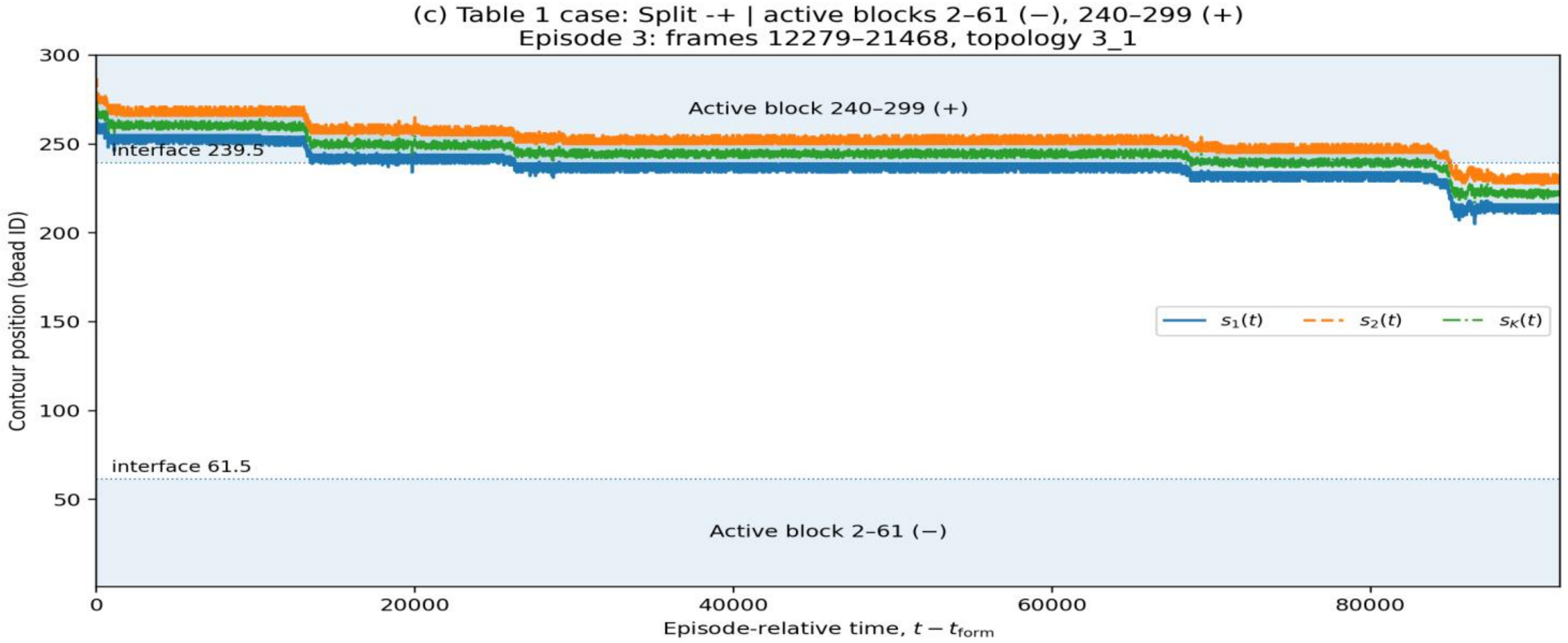

Initial 3_1 segment of the dominant long-lived episode, showing migration toward the passive region and net tightening.

Figure 10

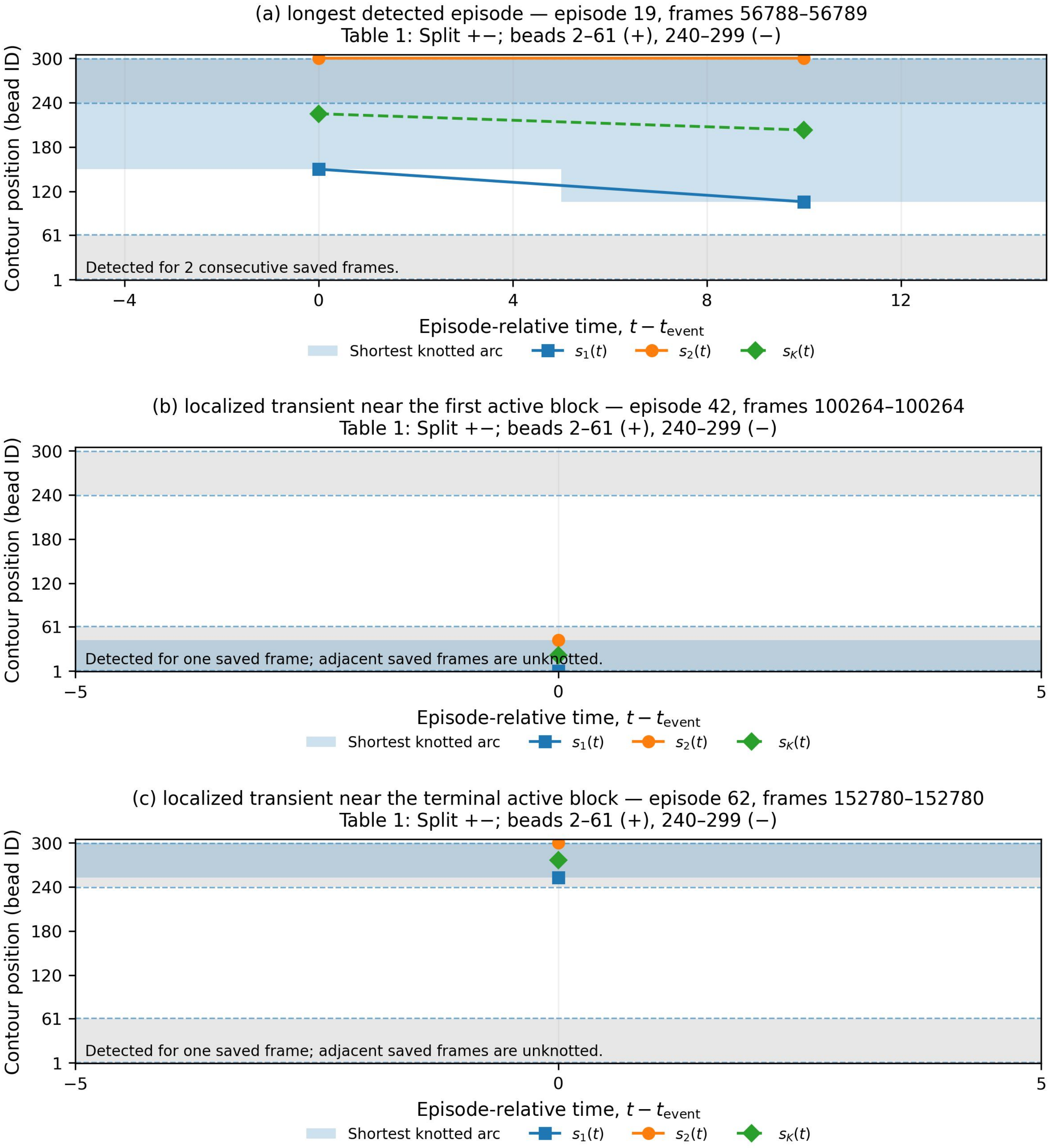

(a) longest detected episode — episode 19, frames 56788–56789
Table 1: Split +−; beads 2–61 (+), 240–299 (−)
Contour position (bead ID)
Detected for 2 consecutive saved frames.
Episode-relative time, $t - t_{\mathrm{event}}$
Shortest knotted arc
$s_1(t)$
$s_2(t)$
$s_K(t)$
(b) localized transient near the first active block — episode 42, frames 100264–100264
Table 1: Split +−; beads 2–61 (+), 240–299 (−)
Contour position (bead ID)
Detected for one saved frame; adjacent saved frames are unknotted.
Episode-relative time, $t - t_{\mathrm{event}}$
Shortest knotted arc
$s_1(t)$
$s_2(t)$
$s_K(t)$
(c) localized transient near the terminal active block — episode 62, frames 152780–152780
Table 1: Split +−; beads 2–61 (+), 240–299 (−)
Contour position (bead ID)
Detected for one saved frame; adjacent saved frames are unknotted.
Episode-relative time, $t - t_{\mathrm{event}}$
Shortest knotted arc
$s_1(t)$
$s_2(t)$
$s_K(t)$

Figure 11

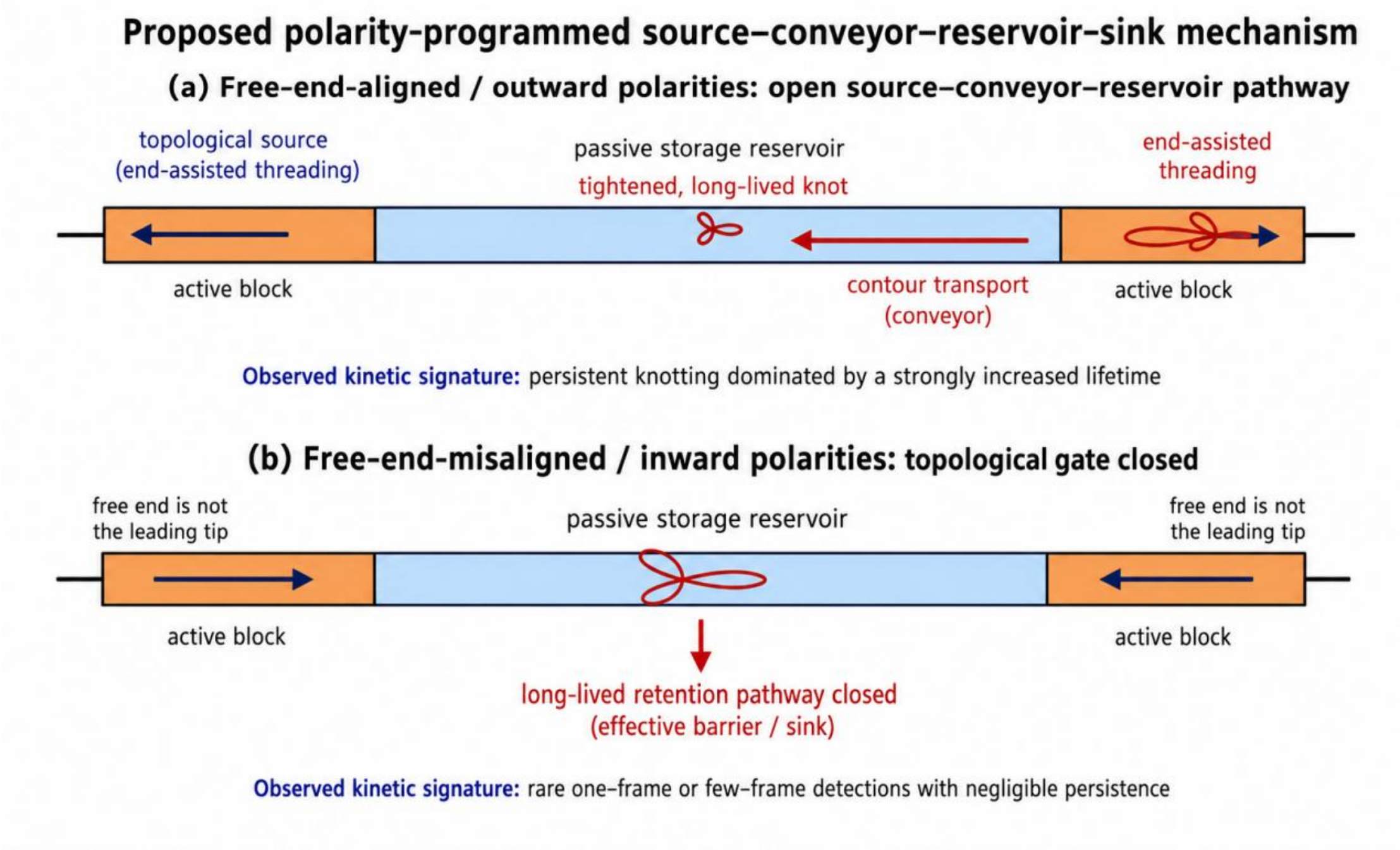
Proposed polarity-programmed source–conveyor–reservoir–sink mechanism
(a) Free-end-aligned / outward polarities: open source–conveyor–reservoir pathway
topological source
(end-assisted threading)
passive storage reservoir
tightened, long-lived knot
end-assisted
threading
active block
contour transport
(conveyor)
active block
Observed kinetic signature: persistent knotting dominated by a strongly increased lifetime
(b) Free-end-misaligned / inward polarities: topological gate closed
free end is not
the leading tip
passive storage reservoir
free end is not
the leading tip
active block
active block
long-lived retention pathway closed
(effective barrier / sink)
Observed kinetic signature: rare one–frame or few–frame detections with negligible persistence

Figure 12

Table 1. Knotting statistics for $N = 300$, $p = 0.4$, and $|Pe| = 6$. The Single + active block occupies beads 180-299. The split architectures use active blocks on beads 2-61 and 240-299; polarity pairs are listed in increasing-bead-index order, with + and - denoting propulsion toward increasing and decreasing bead index, respectively. The observables are the total knotting probability $P_K$, the joint active- and passive-center probabilities $P_{K,\mathrm{act}}$ and $P_{K,\mathrm{pas}}$ with $P_K = P_{K,\mathrm{act}} + P_{K,\mathrm{pas}}$, the dimensionless conditional formation rate $\alpha_K^0 t$, the mean and median knot-center coordinates $\langle i_K \rangle$ and $\mathrm{median}(i_K)$, the mean and median shortest-knotted-arc lengths $\langle L_K \rangle$ and $\mathrm{median}(L_K)$, the mean active-bead fraction $\langle \varphi_K^{\mathrm{act}} \rangle$, and the mean and median completed-episode lifetimes $\frac{\langle \tau_K \rangle}{t}$ and $\frac{\mathrm{median}(\tau_K)}{t}$. Values in parentheses are bootstrap standard errors. Right-censored episodes are excluded from the reported completed-episode lifetimes. For the extremely rare Split +- case, zero or very small errors reflect sparse observations rather than high statistical precision.

| Observable | Single + | Split ++ | Split -- | Split -+ | Split +- |
|---|---|---|---|---|---|
| $P_K$ | 0.6875 (0.02963) | 0.1793 (0.02751) | 0.1786 (0.02631) | 0.9354 (0.01643) | 3.897e-4 (4.762e-5) |
| $P_{K,\mathrm{act}}$ | 0.006695 (0.001616) | 0.001858 (0.0008949) | 0.001872 (0.0007751) | 0.03773 (0.01331) | 6.456e-5 (1.807e-5) |
| $P_{K,\mathrm{pas}}$ | 0.6808 (0.02969) | 0.1775 (0.02744) | 0.1768 (0.02619) | 0.8976 (0.02098) | 3.251e-4 (4.483e-5) |
| $\alpha_K^0 t$ | 1.453e-4 (2.105e-5) | 4.309e-5 (5.096e-6) | 3.727e-5 (5.388e-6) | 1.173e-4 (2.056e-5) | 3.847e-5 (4.613e-6) |
| $\langle i_K \rangle$ | 100.97 (2.760) | 155.9 (6.153) | 148.6 (5.605) | 161.1 (1.450) | 140.4 (7.207) |
| $\mathrm{median}(i_K)$ | 99.73 (3.273) | 153.4 (9.913) | 152.3 (8.268) | 157.1 (0.2079) | 138.9 (12.91) |
| $\langle L_K \rangle$ | 51.83 (2.937) | 35.68 (3.340) | 31.22 (1.846) | 125.4 (2.282) | 188.6 (7.912) |
| $\mathrm{median}(L_K)$ | 34.39 (3.245) | 29.19 (2.623) | 28.76 (1.958) | 139.7 (1.417) | 189.4 (10.34) |
| $\langle \phi_K^{\mathrm{act}} \rangle$ | 0.01687 (0.003387) | 0.01502 (0.006297) | 0.01295 (0.005422) | 0.04019 (0.01241) | 0.4040 (0.01942) |
| $\langle \tau_K \rangle / t$ | 15326.10 (2649.23) | 5069.36 (764.47) | 5856.40 (804.17) | 121952.87 (38384.56) | 10.13 (0.13) |
| $\mathrm{median}(\tau_K)/t$ | 8615.69 (3261.77) | 1530.36 (2264.98) | 5048.08 (3217.72) | 84709.17 (43775.66) | 10.0 (0.0) |

Table 2. Knotting statistics for $N = 300$, $p = 0.3$, and $|Pe| = 6$. The Single + active block occupies beads 210-299. The standard split geometry uses active blocks on beads 2-46 and 180-224; the terminal variant Split -+ T retains the first block and moves the positive second block to beads 255-299. Polarity pairs are listed in increasing-bead-index order, with + and - denoting propulsion toward increasing and decreasing bead index, respectively, and T denoting terminal placement of the second block. The observables are $P_K$, $P_{K,\mathrm{act}}$, $P_{K,\mathrm{pas}}$, $\alpha_K^0 t$, $\langle i_K \rangle$, median$(i_K)$, $\langle L_K \rangle$, median$(L_K)$, $\langle \varphi_K^{\mathrm{act}} \rangle$, $\frac{\langle \tau_K \rangle}{t}$, and $\frac{\mathrm{median}(\tau_K)}{t}$, as defined in Table 1. Values in parentheses are bootstrap standard errors. Right-censored episodes are excluded from the reported completed-episode lifetimes. Zero or very small errors in one-frame rare-event cases indicate limited sampling rather than high precision.

| Observable | Single + | Split ++ | Split -- | Split -+ | Split -+ T | Split +- |
|---|---|---|---|---|---|---|
| $P_K$ | 0.7257 (0.02865) | 1.258e-4 (2.385e-5) | 0.1222 (0.02205) | 5.060e-6 (4.992e-6) | 0.6461 (0.03304) | 5.060e-6 (4.992e-6) |
| $P_{K,\mathrm{act}}$ | 0.007453 (0.004380) | 5.055e-6 (5.020e-6) | 0.02121 (0.007404) | 5.060e-6 (4.992e-6) | 0.2117 (0.02811) | 5.060e-6 (4.992e-6) |
| $P_{K,\mathrm{pas}}$ | 0.7183 (0.02903) | 1.207e-4 (2.348e-5) | 0.1010 (0.01944) | 0.0 (0.0) | 0.4344 (0.03451) | 0.0 (0.0) |
| $\alpha_K^0 t$ | 1.189e-4 (1.433e-5) | 1.210e-5 (2.243e-6) | 2.443e-5 (4.416e-6) | 5.060e-7 (4.992e-7) | 6.697e-5 (5.965e-6) | 5.060e-7 (4.992e-7) |
| $\langle i_K \rangle$ | 116.1 (3.480) | 116.4 (8.551) | 159.4 (12.53) | 28.0 (0.0) | 251.5 (0.3534) | 28.0 (0.0) |
| median$(i_K)$ | 116.4 (5.375) | 123.2 (12.00) | 146.1 (20.56) | 28.0 (0.0) | 250.1 (0.6657) | 28.0 (0.0) |
| $\langle L_K \rangle$ | 55.39 (3.068) | 213.3 (12.62) | 42.94 (6.181) | 57.0 (0.0) | 15.35 (0.0675) | 57.0 (0.0) |
| median$(L_K)$ | 37.97 (3.209) | 234.7 (29.81) | 30.10 (1.591) | 57.0 (0.0) | 15.0 (0.0) | 57.0 (0.0) |
| $\langle \phi_K^{\mathrm{act}} \rangle$ | 0.01360 (0.004407) | 0.3290 (0.01718) | 0.09910 (0.02742) | 0.7895 (5.104e-15) | 0.3133 (0.02264) | 0.7895 (5.104e-15) |
| $\langle \tau_K \rangle / t$ | 22353.34 (3458.22) | 10.40 (0.41) | 5746.25 (942.42) | 10.0 (0.0) | 27266.97 (3354.29) | 10.0 (0.0) |
| median$(\tau_K)/t$ | 14686.54 (4508.66) | 10.0 (0.0) | 5760.60 (3142.90) | 10.0 (0.0) | 20531.94 (3526.86) | 10.0 (0.0) |

[23] S. K. Anand and S. P. Singh, "Conformation and dynamics of a self-avoiding active flexible polymer," Physical Review E 101, 030501(R) (2020). https://doi.org/10.1103/PhysRevE.101.030501.

[24] C. A. Philipps, G. Gompper, and R. G. Winkler, "Tangentially driven active polar linear polymers—An analytical study," The Journal of Chemical Physics 157, 194904 (2022). https://doi.org/10.1063/5.0120493.

[25] D. Loi, S. Mossa, and L. F. Cugliandolo, "Non-conservative forces and effective temperatures in active polymers," Soft Matter 7, 10193-10209 (2011). https://doi.org/10.1039/C1SM05819C.

[26] A. Martín-Gómez, T. Eisenstecken, G. Gompper, and R. G. Winkler, "Hydrodynamics of polymers in an active bath," Physical Review E 101, 052612 (2020). https://doi.org/10.1103/PhysRevE.101.052612.

[27] R. F. Rodríguez, J.R. Gomez-Solano, J. Fujioka, Active diffusion model and dynamic structure factor of self-propelled particles in a three-parameter fluctuating Mittag-Leffler fluid, Physica A 662 (2025) 130411. DOI: 10.1016/j.physa.2025.130411.

[28] H. Yücel, Sorting chiral active Brownian particles with radially modulated optical fields, Physica A 688 (2026) 131411. DOI: 10.1016/j.physa.2026.131411.

[29] M. Foglino, E. Locatelli, C. A. Brackley, D. Michieletto, C. N. Likos, and D. Marenduzzo, "Non-equilibrium effects of molecular motors on polymers," Soft Matter 15, 5995-6005 (2019). https://doi.org/10.1039/C9SM00273A.

[30] J.-X. Li, S. Wu, L.-L. Hao, Q.-L. Lei, and Y.-Q. Ma, "Activity-driven polymer knotting for macromolecular topology engineering," Science Advances 10, eadr0716 (2024). https://doi.org/10.1126/sciadv.adr0716.

[31] M. Vatin, E. Orlandini, and E. Locatelli, "Upsurge of spontaneous knotting in polar diblock active polymers," Physical Review Letters 134, 168301 (2025). https://doi.org/10.1103/PhysRevLett.134.168301.

[32] Z. Zhang, L. Li, Y. Zhu, R. Zhang, M. Yang, and L. Dai, "Behavior of active polymer knots," Macromolecules 58, 11229-11236 (2025). https://doi.org/10.1021/acs.macromol.5c01381.

[33] J. D. Weeks, D. Chandler, and H. C. Andersen, "Role of repulsive forces in determining the equilibrium structure of simple liquids," The Journal of Chemical Physics 54, 5237-5247 (1971). https://doi.org/10.1063/1.1674820.